\documentclass[9pt,a4paper]{article}
\usepackage[margin=0.6in]{geometry}

\renewcommand{\title}[1]{\vbox{\center\LARGE{#1}}\vspace{5mm}}
\renewcommand{\author}[1]{\vbox{\center#1}\vspace{5mm}}
\newcommand{\address}[1]{\vbox{\center\em#1}}
\newcommand{\email}[1]{\vbox{\center\tt#1}\vspace{5mm}}

\renewcommand{\date}[1]{\vbox{\center#1}}

\usepackage{amssymb}
\usepackage{amsmath,bm}
\usepackage{amssymb}
\usepackage{graphicx}
\usepackage{amsfonts}         
\usepackage{fancybox}   

\usepackage{enumitem}

\usepackage{slashed}

\usepackage{epstopdf}
\usepackage[usenames,dvipsnames]{xcolor}
\usepackage[pagebackref, bookmarks={false}, pdfauthor={JD}, pdftitle={aitojibonkalida}]{hyperref}
\hypersetup{colorlinks=true, linkcolor=darkblue, citecolor=red, filecolor=OliveGreen, urlcolor=blue, filebordercolor={.8 .8 1}, urlbordercolor={.8 .8 0}}
\usepackage{soul}
\setstcolor{Red}

\usepackage{wrapfig}

\definecolor{jazzberryjam}{rgb}{0.65, 0.04, 0.37}
\definecolor{lust}{rgb}{0.9, 0.13, 0.13}
\definecolor{sandybrown}{rgb}{0.96, 0.64, 0.38}
\definecolor{mountainmeadow}{rgb}{0.19, 0.73, 0.56}
\definecolor{glaucous}{rgb}{0.38, 0.51, 0.71}
\definecolor{chromeyellow}{rgb}{1.0, 0.65, 0.0}
\definecolor{emerald}{rgb}{0.31, 0.78, 0.47}
\definecolor{deepsaffron}{rgb}{1.0, 0.6, 0.2}
\definecolor{darkgreen}{rgb}{0,0.4,0}
\definecolor{darkred}{rgb}{0.4,0,0}
\definecolor{darkblue}{rgb}{0,0,0.4}
\definecolor{lightblue}{rgb}{.6,.6,0.9}

\definecolor{uglybrown}{rgb}{0.8,  0.7,  0.5}

\definecolor{palatinatepurple}{rgb}{0.41, 0.16, 0.38}
\definecolor{celebrationcolor}{rgb}{0.75,  0.0,  0.9}

 \usepackage{mdframed}

\usepackage{framed}
\definecolor{shadecolor}{rgb}{0.90,0.90,0.90}

\usepackage{tkz-euclide}

\usepackage{makeidx}
\usepackage{cite}
\usepackage{bm}
\usepackage{geometry}
\usepackage{braket}
\usepackage[margin=20pt,small]{caption}

\usepackage[toc]{appendix}

\usepackage{tikz}
\usetikzlibrary{calc,decorations.markings,arrows.meta,shapes.misc,decorations.pathmorphing,calc,bending}

\usetikzlibrary{arrows.meta,shapes.misc,decorations.pathmorphing,calc,bending}

\tikzset{
  branch point/.style={cross out,draw=black,fill=none,minimum size=2*(#1-\pgflinewidth),inner sep=0pt,outer sep=0pt}, 
  branch point/.default=5
}
\tikzset{
  branch cut/.style={
    decorate,decoration=snake,
    to path={
      (\tikztostart) -- (\tikztotarget) \tikztonodes
    },
    }
  }

\input epsf

\newlength{\extraspace}
\newlength{\extraspaces}
\def\be{\begin{equation}}
\def\ee{\end{equation}}

\newcommand{\bea}{\begin{eqnarray}}
\newcommand{\eea}{\end{eqnarray}}

\def\a{\alpha}

\def\II{\relax{I\kern-.10em I}}

\def\IB{\relax{\rm I\kern-.18em B}}

\def\ID{\relax{\rm I\kern-.18em D}}
\def\IE{\relax{\rm I\kern-.18em E}}
\def\IF{\relax{\rm I\kern-.18em F}}
\def\IG{\relax\hbox{$\inbar\kern-.3em{\rm G}$}}
\def\IGa{\relax\hbox{${\rm I}\kern-.18em\Gamma$}}
\def\IH{\relax{\rm I\kern-.18em H}}
\def\II{\relax{\rm I\kern-.18em I}}
\def\IK{\relax{\rm I\kern-.18em K}}

\def\inbar{\,\vrule height1.5ex width.4pt depth0pt}

\def\lp10{\ell_p^{10}}
\def\lp11{\ell_p^{11}}
\def\R11{R_{11}}

\def\frac#1#2{{#1 \over #2}}

\newdimen\tableauside\tableauside=1.0ex
\newdimen\tableaurule\tableaurule=0.4pt
\newdimen\tableaustep
\def\phantomhrule#1{\hbox{\vbox to0pt{\hrule height\tableaurule width#1\vss}}}
\def\phantomvrule#1{\vbox{\hbox to0pt{\vrule width\tableaurule height#1\hss}}}
\def\sqr{\vbox{%
  \phantomhrule\tableaustep
  \hbox{\phantomvrule\tableaustep\kern\tableaustep\phantomvrule\tableaustep}%
  \hbox{\vbox{\phantomhrule\tableauside}\kern-\tableaurule}}}
\def\squares#1{\hbox{\count0=#1\noindent\loop\sqr
  \advance\count0 by-1 \ifnum\count0>0\repeat}}
\def\tableau#1{\vcenter{\offinterlineskip
  \tableaustep=\tableauside\advance\tableaustep by-\tableaurule
  \kern\normallineskip\hbox
    {\kern\normallineskip\vbox
      {\gettableau#1 0 }%
     \kern\normallineskip\kern\tableaurule}%
  \kern\normallineskip\kern\tableaurule}}
\def\gettableau#1 {\ifnum#1=0\let\next=\null\else
  \squares{#1}\let\next=\gettableau\fi\next}

 \def\eqnn#1{\xdef #1{(\secsym\the\meqno)}\writedef{#1\leftbracket#1}%
 \global\advance\meqno by1\wrlabeL#1}
 \def\eqna#1{\xdef #1##1{\hbox{$(\secsym\the\meqno##1)$}}
 \writedef{#1\numbersign1\leftbracket#1{\numbersign1}}%
 \global\advance\meqno by1\wrlabeL{#1$\{\}$}}
 \def\eqn#1#2{\xdef #1{(\secsym\the\meqno)}\writedef{#1\leftbracket#1}%
 \global\advance\meqno by1$$#2\eqno#1\eqlabeL#1$$}

\global\newcount\itemno \global\itemno=0

\def\itemaut#1{\global\advance\itemno by1\noindent\item{\the\itemno.}#1}

\def\({\left(}
\def\){\right)}

\def\lsim{\mathrel{\mathstrut\smash{\ooalign{\raise2.5pt\hbox{$<$}\cr\lower2.5pt\hbox{$\sim$}}}}}
\def\gsim{\mathrel{\mathstrut\smash{\ooalign{\raise2.5pt\hbox{$>$}\cr\lower2.5pt\hbox{$\sim$}}}}}

\def\overleftrightarrow#1{\vbox{\ialign{##\crcr
     $\leftrightarrow$\crcr\noalign{\kern-0pt\nointerlineskip}
     $\hfil\displaystyle{#1}\hfil$\crcr}}}
     
     \def\overleftarrow#1{\vbox{\ialign{##\crcr
     $\leftarrow$\crcr\noalign{\kern-0pt\nointerlineskip}
     $\hfil\displaystyle{#1}\hfil$\crcr}}}

\usepackage[yyyymmdd,hhmmss]{datetime}
\newif{\ifeq}           
\eqtrue                 
\newcounter{lecturecounter}

\usepackage[utf8]{inputenc}
\usepackage{comment}
\usepackage{float, xcolor}
\usepackage{physics}
\usepackage{ragged2e}
\usepackage{booktabs}
\usepackage{latexsym}
\usepackage{mathtools}

\usepackage{subcaption}
\usepackage{graphicx}
\usepackage{cleveref}
\usepackage{amsmath}

\numberwithin{equation}{section}
\definecolor{darkgreen}{rgb}{0,0.5,0}
\definecolor{darkblue}{rgb}{0,0,0.6}

\hypersetup{
citecolor=blue,
colorlinks=true,
filecolor=red,
linkcolor=blue,
linktocpage=true,
urlcolor=darkgreen
}

\begin{document}
\title{Reflected entropy and Markov gap in non-inertial frames}

\author{Jaydeep Kumar Basak${}^{|0\rangle,|1\rangle}$, Dimitrios Giataganas${}^{|0\rangle,|1\rangle,|2\rangle}$, Sayid Mondal${}^{|3\rangle}$ and Wen-Yu Wen${}^{|3\rangle,|4\rangle}$}

\address{ \vspace{0.4cm}
	{\it $|0\rangle$
		Department of Physics,\\ National Sun Yat-Sen University, \\ Kaohsiung 80424, Taiwan\\}
	{\it $|1\rangle$
		Center for Theoretical and Computational Physics, \\ Kaohsiung 80424, Taiwan\\}
	{\it $|2\rangle$
		Physics Division, National Center for Theoretical Sciences,\\Taipei 10617, Taiwan\\}
	{\it $|3\rangle$
		Center for High Energy Physics and Department of Physics,\\Chung-Yuan Christian University, Taoyuan, Taiwan \\}
	{\it $|4\rangle$
		Leung Center for Cosmology and Particle Astrophysics,\\ 	National Taiwan University, Taipei, Taiwan\\
}}

\date{}			

\email{\href{mailto:jkb.hep@gmail.com}{jkb.hep@gmail.com}, \href{mailto:dimitrios.giataganas@mail.nsysu.edu.tw}{dimitrios.giataganas@mail.nsysu.edu.tw},\\ \href{mailto:sayid.mondal@gmail.com}{sayid.mondal@gmail.com},
\href{mailto:wenw@cycu.edu.tw}{wenw@cycu.edu.tw}}

\abstract{We explore the reflected entropy and the Markov gap between two modes of a free fermionic field as observed by accelerating observers. This is done for both bipartite system which is described by Bell state and tripartite systems which are represented by  Werner and Greenberger–Horne–Zeilinger states. The reflected entropy degrades monotonically as a result of the Unruh effect, eventually reaching a non-zero minimum value in the limit of infinite acceleration. Furthermore, we show that the Markov gap exhibits monotonic behavior with regard to acceleration in all three cases. In addition, we suggest a function for reflected entropy which decreases monotonically with decreasing Unruh temperature for all states. Finally, we confirm that the reflected entropy for our system does reduce under the partial tracing of the degrees of freedom for our states.
}

\newpage
\tableofcontents

\vfill\eject

\section{Introduction }
\label{sec_intro}
Entanglement has emerged as a central issue in diverse areas of theoretical and experimental physics from condensed matter physics to quantum theory of gravity. It has served as a resource of several non local observables for quantum information tasks and quantum communications. A large part of entanglement studies, consists of non-relativistic systems. More recently, the understanding of entanglement has been extended in relativistic settings and has been explored in different directions. It is believed to be important from a fundamental point of view and for applications. Various experiments in quantum information theory involve observers with relativistic velocities which demand a rich theoretical understanding of the characteristics of entanglement in non inertial frames. It is known that entanglement between observers in  inertial frame remains constant. On the other hand, when relativistic non-inertial motion is involved the quantum information becomes observer dependent. A simple system to study this phenomenon is to consider the entanglement of a non-interacting massless field from the point of view of an observer who is uniformly accelerated \cite{Fuentes-Schuller:2004iaz}. One may assume in the inertial frame a maximally entangled pure state, whose modes are obtained from a massless scalar field as the solution of the Klein-Gordon equation in Minkowski coordinates. To describe the state from the point of view of non-inertial observers, the massless scalar field should be now considered in Rindler spacetime. A Bogoliubov transformation on the former solution in Minkowski leads to the later one in the Rindler spacetime \cite{walls2008quantum}. An immediate consequence is that a pure state described by inertial observers becomes mixed for the uniformly accelerated observers. Following this approach it has been found that non-inertial observers see a degradation of the entanglement compared to the inertial ones. The studies have been extended to the fermionic systems, \cite{Alsing:2006cj} following a similar methodology with the solution and their transformation of the Dirac equation in different spacetimes, eventually obtaining the same qualitative results.  

The ground state of a given mode for inertial observers becomes a two modes state for accelerated observers, each one corresponding to the field observed in the two causally disconnected Rindler regions. This is due to the fact that now the state is thermal, where an information loss appears for an observer in one of the regions, since he/she needs to trace over the other region \cite{P_C_W_Davies_1975,PhysRevD.14.870}. So far we have reported the results of the various entanglement measures in the system under study. Nevertheless, a more appropriate, richer measure can be used for the investigation of the correlation of these types of mixed states. In the context of quantum information theory, bipartite entanglement has been studied widely to understand the entanglement structure of any system. Several attempts have been made to explore the multipartite entanglement. This type of correlation has a wide range of applications in various quantum phenomena ranging from quantum gravity to quantum computation. Despite its importance, multipartite entanglement measure is still a challenging field of research in quantum information theory (see \cite{Liu:2023pdz,Liu:2021ctk,Zou:2020bly,PhysRevLett.128.176402} and reference therein for recent progress).

More recently,  the so called reflected entropy has been proposed as a crucial tool to investigate the correlation of a mixed state \cite{Dutta:2019gen}. This measure involves a canonical purification of a mixed state which is easier to obtain compared to the purification considered in the computation of the entanglement of purification. The computation of reflected entropy was introduced for a conformal field theory (CFT) based on a specific replica technique. The entanglement wedge cross section has been suggested as the holographic dual of reflected entropy in the framework of AdS/CFT correspondence. Note that entanglement wedge cross section has also been proposed to be dual to the entanglement of purification in \cite{Takayanagi:2017knl,Umemoto:2018jpc}. Furthermore, it was argued that the tripartite entanglement is necessary for holographic CFT states in order to respect the conjectures about the reflected entropy or entanglement of purification involving the entanglement wedge cross section \cite{Akers:2019gcv}. These results indicate that the reflected entropy inherits some information about the multipartite entanglement by studying a two party state \footnote{There are other entanglement measures i.e. three-tangle\cite{Coffman:1999jd}, $\pi$-tangle\cite{PhysRevA.75.062308} in the literature which are used frequently to quantify tripartite entanglement.}. Following these developments, two non negative measures of tripartite entanglement, named as $\emph{g}$ and $\emph{h}$, have been proposed in \cite{Zou:2020bly}. The measure $\emph{g}$ is defined as the difference between the reflected entropy and the mutual information whereas for $\emph{h}$, it is the difference between the double of entanglement of purification and the mutual information. 
Furthermore, the quantity $\emph{g}$ has been explored in \cite{Hayden:2021gno} from an information theoretic point of view where it was related to a specific Markov recovery problem and thus the name Markov gap was coined. A non-vanishing value of the Markov gap precludes a perfect Markov recovery map. It has been also demonstrated that the lower bound of the Markov gap in a holographic CFT state is related to the number of boundaries of the entanglement wedge cross section. Despite of the success of the reflected entropy in the context of AdS/CFT duality, the monotonicity of this measure under partial tracing, which is a requirement of a good measure of correlation, has been questioned very recently in \cite{Hayden:2023yij}. For a qutrit-qutrit-qubit system, the density matrix of a fine-tuned quantum state can violate the monotonicity of the reflected entropy for the Renyi index $\xi\in (0,2)$.

These developments generate an intense interest to understand the reflected entropy from the viewpoint of quantum information theory. In this article we extend these studies on fermionic systems in non-inertial frames. The two leading protagonists are the reflected entropy and the Markov gap considered for three different scenarios. In the first case, we have two observers, one stationary (Alice) and the other accelerating (Bob), who shared a bipartite entangled fermionic mode described by the Bell state in an inertial frame. In the second and third scenarios, there are three observers, with Alice and Charlie being stationary and Bob accelerating uniformly, who initially shared a tripartite entangled fermionic mode described by the Werner state (W-state) and Greenberger–Horne–Zeilinger (GHZ) state. We study in detail the reflected entropy for these states. To begin with, we show that reflected entropy is monotonic under partial trace for our states which indicates that it is a good measure of correlation at least for the states in question and the acceleration of the observers we consider.  Reflecting on the recent developments this is a necessary check. As a side relevant exercise we show that exist new states (with no acceleration involved) in higher dimensional Hilbert spaces that violate the monotonicity of reflected entropy confirming and extending the work of \cite{Hayden:2023yij}. Getting back to our system we study the properties of the reflected entropy for all our states. We find a degradation of correlation between Alice and Bob due to the Unruh effect in all three scenarios. In the limit of infinite acceleration, the reflected entropy reaches to a non-zero minimum value. Meanwhile, the Markov gap between Alice and Bob exhibits a monotonic behavior with respect to acceleration and we notice that it increases for the Bell and GHZ states whereas it decreases with acceleration for the W-state. Furthermore, we have defined a specific dimensionless function, which we call a $\sigma$-function, that depends on reflected entropy which, in all scenarios, exhibits monotonic behavior with Unruh temperature and shows interesting properties.

This paper is arranged as follows: in \cref{set_up} we explain the setup, defining the states and the effect of acceleration on them. These are the states we study later in this article. In the next \cref{ref_en} we present the results for reflected entropy and also study its bounds in the non-inertial frames. In \cref{MG}, we analyze the Markov gap which indicates a specific evolution of three party correlation. Next in \cref{c_function}, we discuss a monotonic $\sigma$-function  based on reflected entropy. Finally, in section \ref{sum_dis}, we summarize our results and present some of the future directions of our work. Our results of the main text are supported by three appendices.

\section{The States and the Non-inertial Observers}\label{set_up}
We consider a free Dirac field in $(1+1)$-dimensional Minkowski space with coordinates $x^\mu=(t,z)$
\begin{equation}
	i \gamma^\mu \partial_\mu \psi-m \psi=0~,
\end{equation}
where $m$ is the particle mass, $\psi$ is the spinor wave function and $\gamma^{\mu}$ are the Dirac gamma matrices. This field may be expanded in terms of positive (fermions) $\psi_k^+$ and negative (anti-fermions)  $\psi_k^-$ energy solutions as
\begin{equation}\label{mode_expan}
	\psi=\int d k\left(a_k \psi_k^{+}+b_k^{\dagger} \psi_k^{-}\right)~,
\end{equation}
where $k$ is the momentum. The Minkowski creation and annihilation operators $(a_k^{\dagger}, b_k^{\dagger})$ and $(a_k, b_k)$
for fermions and anti-fermions satisfy the anticommutation relations
\begin{equation}\label{coomutation_rel}
	\left\{a_i, a_j^{\dagger}\right\}=\left\{b_i, b_j^{\dagger}\right\}=\delta_{i j}~,
\end{equation}
with all other anticommutators vanishing. The Minkowski vacuum state is given as
\begin{equation}\label{Min_vac}
	|0\rangle=\prod_{k k^{\prime}}\left|0_k\right\rangle^{+}\left|0_{k^{\prime}}\right\rangle^{-},
\end{equation}
where the $\{+,-\}$ superscript on the kets indicates the fermion and anti-fermion vacua. Note that as $(a_k^{\dagger})^2=(b_k^{\dagger})^2=0$, there are only two allowed states for each mode, $\left|0_k\right\rangle^{+}$and $\left|1_k\right\rangle^{+}=a_k^{\dagger}\left|0_k\right\rangle^{+}$ for fermions, and $\left|0_k\right\rangle^{-}$and $\left|1_k\right\rangle^{-}=b_k^{\dagger}\left|0_k\right\rangle^{-}$ for anti-fermions.

In our work, we  consider three distinct scenarios. In the first case, we consider two non-inertial observers sharing an initially entangled bipartite fermionic field modes described by the Bell state which is given as\footnote{From now on, we will only consider the fermionic field modes and we will also omit the superscript $\{+\}$ and subscript $k$ on the kets.}
\begin{equation}\label{weighted_bell_state}
	\left|B\right\rangle_{AB}=\alpha\left|0\right\rangle_{A}\left|0\right\rangle_{B}+ \sqrt{1-\alpha^2}\left|1\right\rangle_{A}\left|1\right\rangle_{B},\quad \alpha \in(0,1)~,
\end{equation}
where the subscripts $A$ and $B$ indicate the modes associated with the observers Alice and Bob respectively. In the second and third case we consider two tripartite entangled fermionic field modes  represented by the Werner and GHZ states which are given as
\begin{equation}\label{weighted_werner_state}
	\left|W\right\rangle_{ABC}=\alpha\left|1\right\rangle_{A}\left|0\right\rangle_{B}\left|0\right\rangle_{C}+\alpha\left|0\right\rangle_{A}\left|0\right\rangle_{B}\left|1\right\rangle_{C}+ \sqrt{1-2 \alpha^2}\left|0\right\rangle_{A}\left|1\right\rangle_{B}\left|0\right\rangle_{C},\quad \alpha \in(0,\frac{1}{\sqrt{2}})~,
\end{equation}
and
\begin{equation}\label{weighted_GHZ_state}
	\left|GHZ\right\rangle_{ABC}=\alpha\left|0\right\rangle_{A}\left|0\right\rangle_{B}\left|0\right\rangle_{C}+ \sqrt{1-\alpha^2}\left|1\right\rangle_{A}\left|1\right\rangle_{B}\left|1\right\rangle_{C}, \quad \alpha \in(0,1)~,
\end{equation}
where   the subscripts $A$, $B$ and $C$ indicate the modes associated with the observers Alice, Bob and Charlie.

At this stage we need to choose which of the observers is stationary and which is accelerating. For the case of  bipartite state \cref{weighted_bell_state}, we choose the observer Alice to be stationary carrying a detector sensitive only to mode $|n\rangle_A$ and  Bob moves with uniform acceleration possessing a  detector that only detects mode $|n\rangle_B$. As for the tripartite states \cref{weighted_werner_state,weighted_GHZ_state}, we choose Alice and Charlie who detect mode $|n\rangle_A$ and mode $|n\rangle_C$ respectively to be stationary, and the accelerating Bob  who detects mode $|n\rangle_B$.

Rindler coordinates $(\tau, \xi)$ are appropriate to describe  an observer moving with uniform acceleration in an inertial plane described by  Minkowski coordinates $(t,z)$. To describe the entire Minkowski space, two different sets of Rindler coordinates are required which differ from each other by an overall change in sign. These sets of coordinates define two causally disconnected Rindler regions $I$ and $II$ that are defined as 
\begin{equation}
	\begin{aligned}
		& t=a^{-1} e^{a \xi} \sinh a \tau, ~~\quad z=a^{-1} e^{a \xi} \cosh a \tau~, \quad \text{region I}~, \\
		& t=-a^{-1} e^{a \xi} \sinh a \tau, \quad z=a^{-1} e^{a \xi} \cosh a \tau~,  \quad \text{region II}~,
	\end{aligned}
\end{equation}
where  $a$ denotes the proper acceleration of the observer Bob. The Rindler regions $I$ and $II$ are causally disconnected,
 the accelerating observer in either region has no access to the other which leads to detection of a thermal mixed state. Henceforth, we will refer the observer in region $I$ as Bob (B) and the observer in region $II$ as anti-Bob ($\bar{B}$).

The Minkowski  and Rindler creation and annihilation operators are related to each other through the Bogoliubov transformation as \cite{PhysRevD.22.1935,greiner1985quantum,10.1143/PTP.88.1,PhysRevD.43.3979,mcmahon2006dirac,Alsing:2006cj}
    \begin{equation}\label{bogoliubov_trans}
    	\left[\begin{array}{c}
    		a_k \\
    		b_{-k}^{\dagger}
    	\end{array}\right]=\left[\begin{array}{cc}
    		\cos r & -e^{-i \phi} \sin r \\
    		e^{i \phi} \sin r & \cos r
    	\end{array}\right]\left[\begin{array}{c}
    		c_k^I \\
    		d_{-k}^{I I \dagger}
    	\end{array}\right],
    \end{equation}
  where   $\left(c_k^I,d_k^I \right)$ and $(c_k^{I^{\dagger}}, d_k^{I \dagger})$ are annihilation and creation operators for fermion and anti-fermion respectively in Rindler region $I$. In  \cref{bogoliubov_trans}, $r=\tan^{-1}\exp(-\frac{\pi \omega }{a})$ is the acceleration parameter ranging from $0 \leqslant r<\pi / 4$ corresponding to $0 \leqslant a<\infty$, and $\omega$  indicates the Rindler mode frequency as measured by the observer Bob with proper acceleration $a$. The phase $\phi$ in \cref{bogoliubov_trans} is unimportant and it can be absorbed in the definition of operators. The corresponding annihilation and creation operators in region $II$ are $(c_k^{I I}, c_k^{I I \dagger})$ and $(d_k^{I I}, d_k^{I I \dagger})$ respectively. Similarly, the Bogoliubov transformation that mixes an anti-fermion modes in region $I$ to fermion modes in region $II$ is given as follows 
  \begin{equation}
  	\left[\begin{array}{c}
  		b_k \\
  		a_{-k}^{\dagger}
  	\end{array}\right]=\left[\begin{array}{cc}
  		\cos r & e^{-i \phi} \sin r \\
  		-e^{-i \phi} \sin r & \cos r
  	\end{array}\right]\left[\begin{array}{c}
  		d_k^I \\
  		c_{-k}^{I I \dagger}
  	\end{array}\right].
  \end{equation}
   By quantizing the fermionic field in the  Minkowski and Rindler frames, respectively, one can relate the  Minkowski particle vacuum for Bob's modes in terms of Rindler Fock states through the Bogoliubov transformations  as \cite{walls2008quantum,Alsing:2006cj}\footnote{Note that, we have employed the single mode approximation as described in \cite{Alsing:2006cj}.},
\begin{equation}\label{two_mode_ss_vacuum}
	\left|0\right\rangle_{B}=\cos r\left|0\right\rangle_B\left|0\right\rangle_{\bar{B}}+ \sin r\left|1\right\rangle_B\left|1\right\rangle_{\bar{B}},
\end{equation}
and the excited state $\left|1\right\rangle_{B}$ is given as 
\begin{equation}\label{two_mode_ss_first}
	\begin{aligned}
		\left|1\right\rangle_{B}=\left|1\right\rangle {}_B\left|0\right\rangle_{\bar{B}}.
	\end{aligned}
\end{equation}
Note that as Bob accelerates through the Minkowski vacuum $|0\rangle$, his detector detects a number of particle given by 
\begin{equation}
\langle 0|c_k^{I \dagger} c_k^I| 0\rangle_B=\frac{1}{1+e^{\hbar \omega / (k_B T)}}~,
\end{equation}
 where the Unruh temperature $T$ is related to the proper acceleration $a$ as
 \begin{equation}
 	T=\frac{a}{ 2 \pi }~.
 \end{equation}

\subsection{Bell state}\label{density_matrix_bell}
The bipartite fermionic field modes described by Bell state \eqref{weighted_bell_state} may be expressed by employing eqs. \eqref{two_mode_ss_vacuum} and \eqref{two_mode_ss_first} as
\begin{equation}\label{weighted_Bell_acc}
	|B\rangle_{AB\bar{B}}=\alpha \cos r |000\rangle_{AB\bar{B}}+\alpha \sin r |011\rangle_{AB\bar{B}}+\sqrt{1-\alpha^2}|110\rangle_{AB\bar{B}},
\end{equation}
where we have denoted $|l\rangle_A|m\rangle_B |n\rangle_{\bar{B}}=|lmn\rangle_{AB\bar{B}}$ and for simplicity, henceforth we will denote  $|lmn\rangle_{AB\bar{B}}$ as $|lmn\rangle_{}$. The mixed density matrices for Alice-Bob $(AB)$, Alice-anti-Bob $(A\bar{B})$ and Bob-anti-Bob  $(B\bar{B})$ are given as follows
\begin{equation}
	\begin{aligned}
		\rho_{AB}^{(B)}=&\alpha^2\cos^2 r|00\rangle\langle 00|+\alpha \sqrt{1-\alpha^2}\cos r(|00\rangle\langle 11|+|11\rangle\langle 00|)+\alpha^2 \sin^2 r |01\rangle\langle 01|+(1-\alpha^2)|11\rangle\langle 11|~,\\
		\rho_{A\bar{B}}^{(B)}=&\alpha^2\cos^2 r|00\rangle\langle 00|+\alpha^2 \sin^2 r |01\rangle\langle 01| +\alpha \sqrt{1-\alpha^2}\sin r(|01\rangle\langle 10|+|10\rangle\langle 01|)+(1-\alpha^2)|10\rangle\langle 10|~,\\
		\rho_{B\bar{B}}^{(B)}=&\alpha^2\cos^2 r|00\rangle\langle 00|+\alpha^2 \cos r \sin r (|00\rangle\langle 11|+|11\rangle\langle 00|)+(1-\alpha^2)|10\rangle\langle 10|+\alpha^2 \sin^2 r |11\rangle\langle 11|~,
	\end{aligned}
\end{equation}
where the superscript refers to the state, in this case the Bell.
Similarly, the density matrices for Alice, Bob and anti-Bob respectively are $\rho_{A}^{(B)}$, $\rho_{B}^{(B)}$ and  $\rho_{\bar{B}}^{(B)}$. They can be found as
\begin{equation}\label{rhoABell}
	\begin{aligned}
		\rho_A^{(B)}=&\alpha^2|0\rangle\langle 0|+\left(1-\alpha^2\right)| 1\rangle\langle 1|~,\\
		\rho_B^{(B)}=&\alpha^2 \cos ^2 r|0\rangle\langle 0|+\left(1-\alpha^2 \cos ^2 r\right)| 1\rangle\langle 1|~,\\
		\rho_{\bar{B}}^{(B)}=& \big(1-\alpha^2 \sin^2 r\big)|0\rangle\langle 0|+\alpha^2 \sin^2 r |1\rangle\langle 1|~.
	\end{aligned}
\end{equation}
\subsection{Werner state}
The tripartite entangled fermionic mode  described by W-state in eq. \eqref{weighted_werner_state} may be express as follows by employing eqs. \eqref{two_mode_ss_vacuum} and \eqref{two_mode_ss_first}
\begin{equation}\label{W-state_2}
	\begin{aligned}
		|W\rangle_{ABC}=&\alpha \cos r |1000\rangle_{AB\bar{B}C}+\alpha \sin r |1110\rangle_{AB\bar{B}C}+\alpha \cos r|0001\rangle_{AB\bar{B}C}+
		\alpha \sin r |0111\rangle_{AB\bar{B}C}+\sqrt{1-2\alpha^2}|0100\rangle_{AB\bar{B}C}.
	\end{aligned}
\end{equation}
The density matrices of $AB$, $A\bar{B}$, and $B\bar{B}$  are given as
\begin{equation}
	\begin{aligned}
		\rho_{AB}^{(W)}= &\alpha^2 \cos^2 r|00\rangle\langle 00|+\left( (1-2\alpha^2)+\alpha^2 \sin^2 r\right)|01\rangle\langle01|+\alpha \sqrt{1-2\alpha^2}\cos r (|10\rangle\langle 01|+|01\rangle\langle 10|)\\
		&+\alpha^2 \cos^2 r|10\rangle\langle 10|+\alpha^2 \sin^2 r |11\rangle\langle 11|~,\\
		\rho_{A\bar{B}}^{(W)}=&((1-2\alpha^2)+\alpha^2 \cos^2 r)|00\rangle\langle 00|+\alpha^2 \sin^2 r |01\rangle\langle 01|+\alpha \sqrt{1-2\alpha^2}\sin r(|00\rangle\langle 11|+|11\rangle\langle 00|)|\\
		&+ \alpha^2 \cos^2 r |10\rangle\langle 10|+\alpha^2 \sin^2 r |11\rangle\langle 11|~,\\
		\rho_{B\bar{B}}^{(W)}= & 2\alpha^2\cos^2 r|00\rangle\langle 00|+2\alpha^2 \cos r \sin r (|00\rangle\langle 11|+|11\rangle\langle 00|)+2\alpha^2 \sin^2 r |11\rangle\langle 11|+ 	(1-2\alpha^2) |10\rangle\langle 10|~.
	\end{aligned}
\end{equation}
While the density matrices of $A$, $B$ and $\bar{B}$ are
\begin{equation}
	\begin{aligned}
		\rho_{A}^{(W)}=&(1-\alpha^2)|0\rangle\langle 0|+\alpha^2|1\rangle\langle 1|~,\\
		\rho_{B}^{(W)}=&2 \alpha ^2 \cos ^2r|0\rangle\langle 0|+(1-2\alpha ^2\cos^2r)|1\rangle\langle 1|~,\\
		\rho_{\bar{B}}^{(W)}=&(1-2\alpha^2 \sin^2 r) |0\rangle\langle0|+2 \alpha^2 \sin^2 r |1\rangle\langle 1|~.
	\end{aligned}
\end{equation}

\subsection{Greenberger–Horne–Zeilinger state}
By employing eqs. \eqref{two_mode_ss_vacuum} and \eqref{two_mode_ss_first}, the GHZ state \cref{weighted_GHZ_state} may further be expressed as
\begin{equation}\label{GHZ-state_2}
	\begin{aligned}
		|GHZ\rangle_{ABC}=&\alpha \cos r |0000\rangle_{AB\bar{B}C}+\alpha \sin r |0110\rangle_{AB\bar{B}C}+\sqrt{1-\alpha^2}|1101\rangle_{AB\bar{B}C}~.
	\end{aligned}
\end{equation}
The density matrices of $AB$, $A\bar{B}$, and $B\bar{B}$ are as follows
\begin{equation}
	\begin{aligned}
		\rho_{AB}^{(GHZ)}= &\alpha^2 \cos^2 r|00\rangle\langle 00|+\alpha^2 \sin^2 r |01\rangle\langle 01|+(1-\alpha^2)|11\rangle\langle 11|~,\\
		\rho_{A\bar{B}}^{(GHZ)}=&\alpha^2 \cos^2 r|00\rangle\langle 00|+\alpha^2 \sin^2 r |01\rangle\langle 01|+(1-\alpha^2)|10\rangle\langle 10|~,\\
				\rho_{B\bar{B}}^{(GHZ)}= & \alpha^2 \cos^2 r|00\rangle\langle 00|+\alpha^2 \cos r \sin r  (|00\rangle\langle 11|+|11\rangle\langle 00|)+(1-\alpha^2)|10\rangle\langle 10|+\alpha^2 \sin^2 r |11\rangle\langle 11|~,
	\end{aligned}
\end{equation}
while the density matrices of $A$, $B$ and $\bar{B}$ read
\begin{equation}
	\begin{aligned}
		\rho_{A}^{(GHZ)}=&\alpha^2|0\rangle\langle 0|+(1-\alpha^2)|1\rangle\langle 1|~,\\
		\rho_{B}^{(GHZ)}=&\alpha ^2 \cos ^2r|0\rangle\langle 0|+(1-\alpha^2 \cos^2 r)|1\rangle\langle 1|~,\\
		\rho_{\bar{B}}^{(GHZ)}=&(1-\alpha^2 \sin^2 r) |0\rangle\langle0|+ \alpha^2 \sin^2 r |1\rangle\langle 1|~.
	\end{aligned}
\end{equation}

\section{Reflected entropy }\label{ref_en}

In this section we study the reflected entropy between the observers Alice-Bob $(AB)$, Alice-anti-Bob $(A\bar{B})$ and Bob-anti-Bob $(B\bar{B})$ for Bell state given by \cref{weighted_Bell_acc}, W-state given by \cref{W-state_2} and GHZ-state given by \cref{GHZ-state_2} respectively. Before delving into the details of the computation we briefly review reflected entropy in quantum information theory. 
 To begin with, we consider a bipartite density matrix $\rho_{AB}$ in a Hilbert space $\mathcal{H}_A \otimes \mathcal{H}_B$, where $\mathcal{H}_A $ and $\mathcal{H}_B $ are Hilbert spaces associated to the subsystems $A$ and $B$ respectively. The entanglement entropy of the subsystem $A$ is defined as the von Neumann entropy of the reduced density matrix $\rho_A=\mathrm{Tr}_B\rho_{AB}$ as,
\begin{equation}
	S(A)=-\operatorname{Tr}\left(\rho_A \log \rho_A\right)~.
\end{equation}
The mutual information which measures the total correlation between the subsystems $A$ and $B$ is defined as
\begin{equation}
	I(A: B)=S(A)+S(B)-S{(A B)}~,
\end{equation}
which is symmetric in $A$ and $B$. As it has been mentioned in the introduction,  for mixed states the entanglement entropy is not the most appropriate entanglement measure and other mixed state entanglement measures are to be used.  
Note that any mixed state $\rho_{AB}$ in quantum information theory may be expressed as a sum of pure states
\begin{equation}\label{densityAB}
	\rho_{A B}=\sum_a p_a \rho_{A B}^{(a)}~,\quad \rho_{A B}^{(a)}=\left|\phi_a\right\rangle\left\langle\phi_a\right|,
\end{equation}
where  $\left|\phi_a\right\rangle$ is an orthonormal basis of $\mathcal{H}_A \otimes \mathcal{H}_B$, and eigenvalues $p_a$ are non-negative $0 \leq p_a \leq 1$. We construct the  Schmidt decomposition of $|\phi_a\rangle$ by choosing appropriate bases $|i_a\rangle_A \in \mathcal{H}_A$ and $|i_b\rangle_B \in \mathcal{H}_b$ as
\begin{equation}\label{Schmidt_decomp_AB}
	\left|\phi_a\right\rangle=\sum_i \sqrt{l_a^i}\left|i_a\right\rangle_A\left|i_a\right\rangle_B~,
\end{equation}
where $l_a^i$ is a non-negative quantity with the normalization $\sum_i l_a^i=1$. By using eq. \eqref{Schmidt_decomp_AB}, the density matrix eq. \eqref{densityAB} may be expressed as
\begin{equation}
	\rho_{A B}=\sum_{a, i, j} p_a \sqrt{l_a^i l_a^j}|i_a\rangle_A|i_a\rangle_B\langle j_a|_A \langle j_a|_B ~.
\end{equation}
We now interpret $\langle j_a|_A $  and $\langle\ j_a|_B$  as states  $ |j_a\rangle_{A^{\star}}$  and  $|j_a\rangle_{B^{\star}}$ on Hilbert spaces $\mathcal{H}_A^{\star}$ and $\mathcal{H}_B^{\star}$ respectively, and define a pure state $\left|\sqrt{\rho_{A B}}\right\rangle \in \mathcal{H}_A \otimes \mathcal{H}_B \otimes \mathcal{H}_A^{\star} \otimes \mathcal{H}_B^{\star}$ as
\begin{equation}\label{reflected_state}
	|\sqrt{\rho_{AB}}\rangle=\sum_{a, i, j}  \sqrt{p_a l_a^i l_a^j}|i_a\rangle_A|i_a\rangle_B |j_a\rangle_{A^{\star}} |j_a\rangle_{B^{\star}}.
\end{equation}
This state $|\sqrt{\rho_{AB}}\rangle$ known as purification of the state $\rho_{AB}$. The reflected entropy between $A$ and $B$ for $\rho_{AB}$ is defined as the von Neumann entropy of $\rho_{AA^{\star}}=\operatorname{Tr}_{B B^{\star}}\left|\sqrt{\rho_{A B}}\right\rangle\left\langle\sqrt{\rho_{A B}}\right|$ which is given as \cite{Dutta:2019gen,Kusuki:2019zsp,Jeong:2019xdr,Kusuki:2019rbk,Kusuki:2019evw}

\begin{equation}\label{reflected_entropy}
	S_R(A: B)=-\operatorname{Tr}_{A A^{\star}}\left[\rho_{A A^{\star}} \log \rho_{A A^{\star}}\right]~.
\end{equation}
It is interesting to note that the reflected entropy is upper bounded by $\text{min}\{2 S_A, 2 S_B\}$ and lower bounded by the mutual information $I(A:B)$ as
\begin{equation}\label{SR_bound}
	\min \{2 S_A, 2 S_B\} \geq S_R(A: B) \geq I(A: B)~.
\end{equation}
For any tripartite pure state reflected entropy satisfies the polygamy inequality which is given as
\begin{equation}\label{polygamy_inequality}
	S_R(A: B)+S_R(A: C) \geq S_R(A: B C)~.
\end{equation}
Apart from these, reflected entropy can also distinguish isospectral density matrices\cite{PhysRevLett.86.5184}. One example of such density matrices are
\begin{equation}\label{isospectral}
	\rho_1=\frac{1}{3}\left(
	\begin{array}{cccc}
		1 & 0 & 0 & 0 \\
		0 & 1 & 1 & 0 \\
		0 & 1 & 1 & 0 \\
		0 & 0 & 0 & 0 \\
	\end{array}
	\right)~,~~~~~~~
	\rho_2=\frac{1}{3}\left(
	\begin{array}{cccc}
		1 & 0 & 0 & 0 \\
		0 & 0 & 0 & 0 \\
		0 & 0 & 0 & 0 \\
		0 & 0 & 0 & 2 \\
	\end{array}
	\right)~,
\end{equation}
where, $\rho_1$ and $\rho_2$ can be written in the basis $\{|00\rangle,|01\rangle,|10\rangle,|11\rangle\}$ by tracing out one party from a W-state, $\left|W\right\rangle_{ABC}=\frac{1}{\sqrt{3}}(\left|100\right\rangle+\left|010\right\rangle+\left|001\right\rangle)$ and GHZ state, $\left|GHZ\right\rangle_{ABC}=\frac{1}{\sqrt{3}}(\left|000\right\rangle+\sqrt{2}\left|111\right\rangle)$ respectively. For example, in this case one can compute $S_R(\rho_1)=1.49$ and $S_R(\rho_2)=.92$ which clearly distinguishes these two isospectral density matrices.

We now turn to the computation of reflected entropy for the bipartite and tripartite fermionic field mode as described in \cref{weighted_bell_state,weighted_werner_state,weighted_GHZ_state}. To compute reflected entropy $S_R(A:B)$, $S_R(A:\bar{B})$, and  $S_R(B:\bar{B})$ between $AB$, $A\bar{B}$, and $B\bar{B}$, we first construct the canonically purified states $|\sqrt{\rho_{AB}}\rangle$,   $|\sqrt{\rho_{A\bar{B}}}\rangle$, and  $|\sqrt{\rho_{B\bar{B}}}\rangle$ by doubling the Hilbert space as mentioned in eq. \eqref{reflected_state}. Now the reflected entropy $S_R(A:B)$, $S_R(A:\bar{B})$, and  $S_R(B:\bar{B})$ are obtained by using the eq. \eqref{reflected_entropy} for  Bell, W and GHZ states (see  \cref{Bell_state} for details).

\begin{figure}[h]
	\centering
	\begin{subfigure}{.45\textwidth}
		\centering
		\includegraphics[width=.9\linewidth]{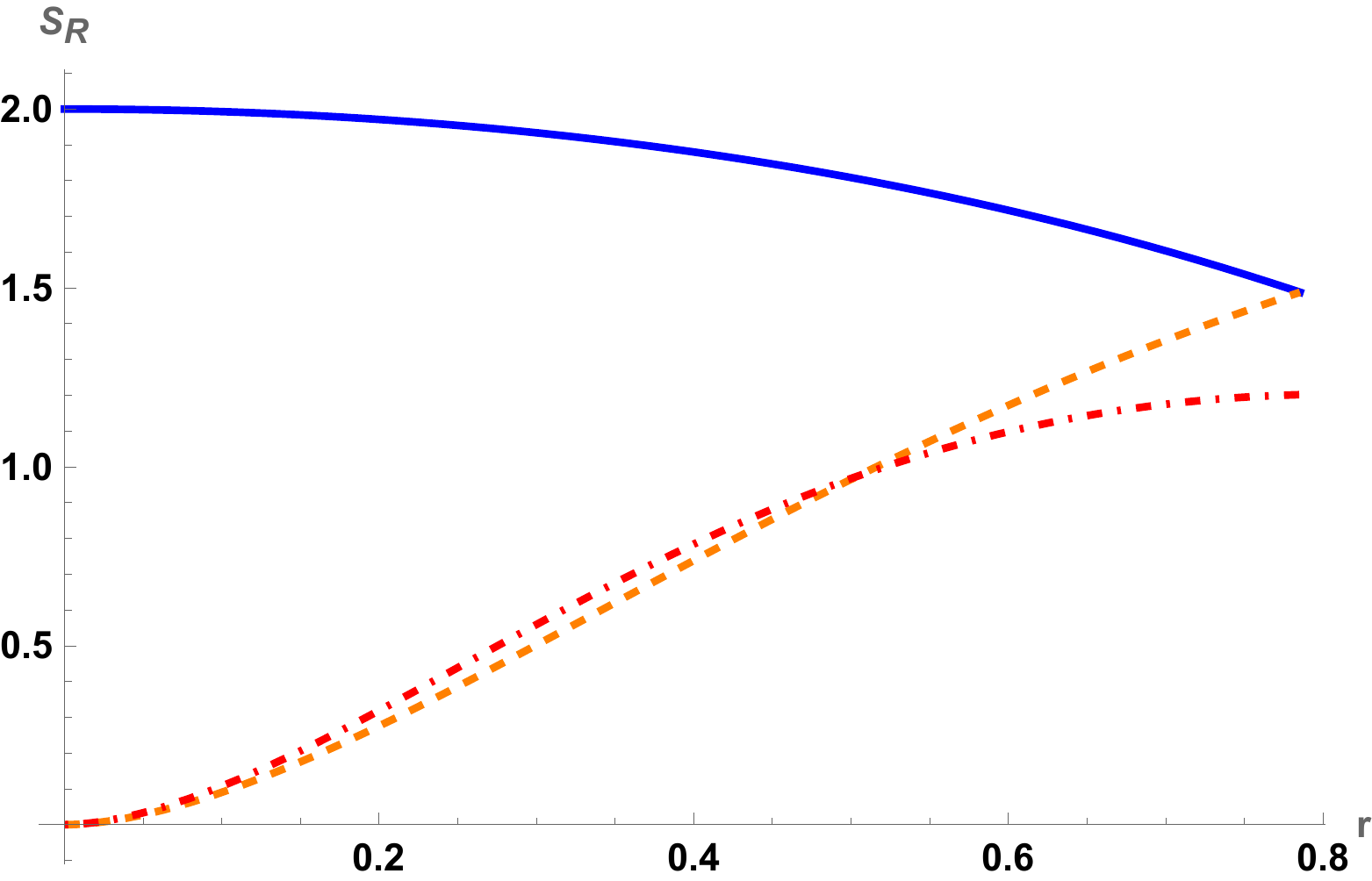}
		\caption{Bell state: $S_R(A:B)$ (Blue solid curve),  $S_R(A:\bar{B})$ (orange dashed curve), and  $S_R(B:\bar{B})$ (red dot-dashed curve). In all  plots that follow for Bell state, $\a$ is fixed to its maximally entangled value for the state unless otherwise stated. 
		}
		\label{fig:Bell_SR_2d}
	\end{subfigure}\hspace{.5cm}
	\begin{subfigure}{.45\textwidth}
		\centering
		\includegraphics[width=.9\linewidth]{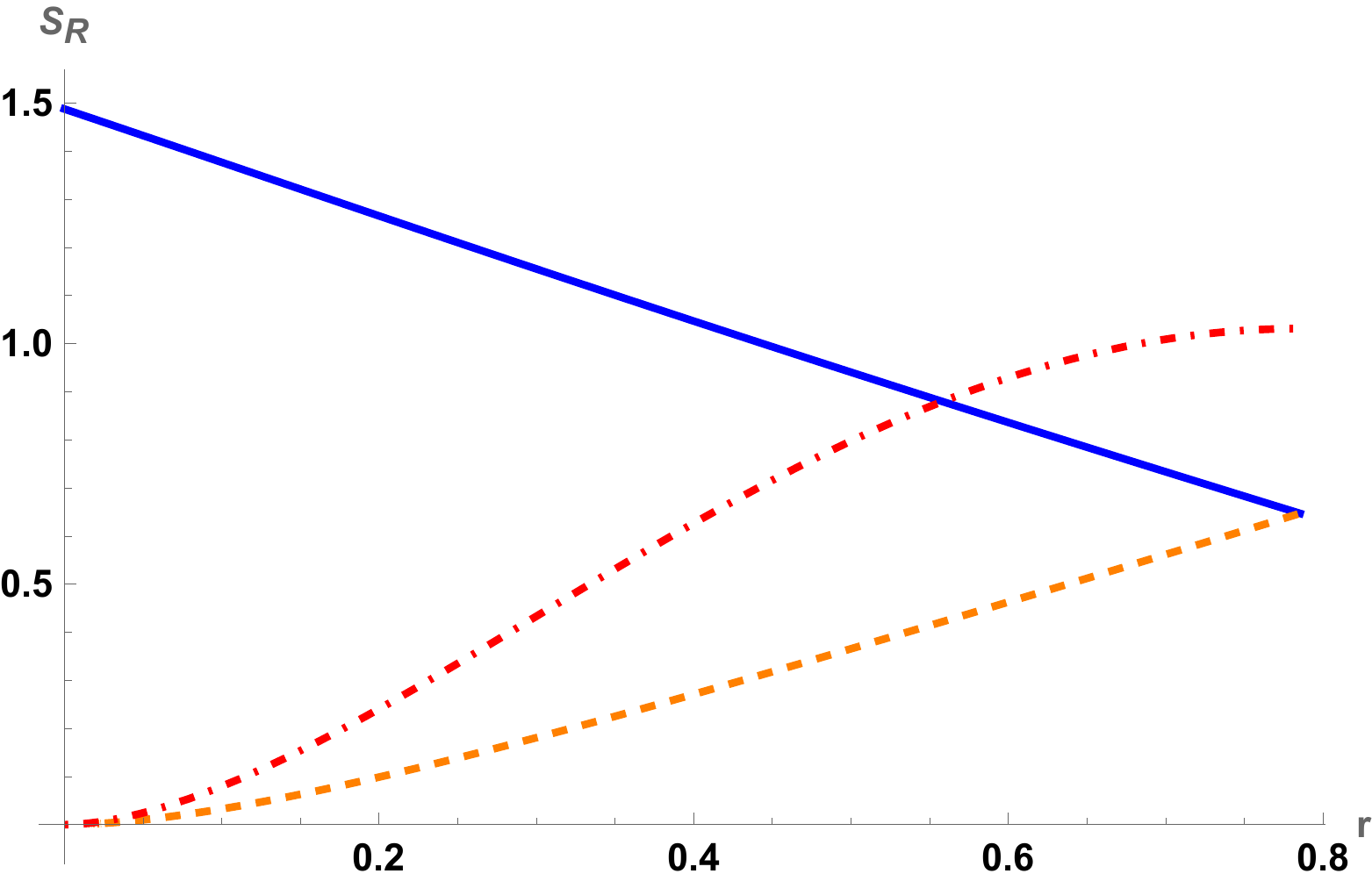}
		\caption{W-state: $S_R(A:B)$ (Blue solid curve),  $S_R(A:\bar{B})$ (orange dashed curve), and  $S_R(B:\bar{B})$ (red dot-dashed curve).  In all  plots that follow for W-state, $\a$ is fixed to its maximally entangled value for the state unless otherwise stated.
		}
		\label{fig:GHZ_SR_2d}
	\end{subfigure}
\begin{subfigure}{.45\textwidth}
	\centering
	\includegraphics[width=.9\linewidth]{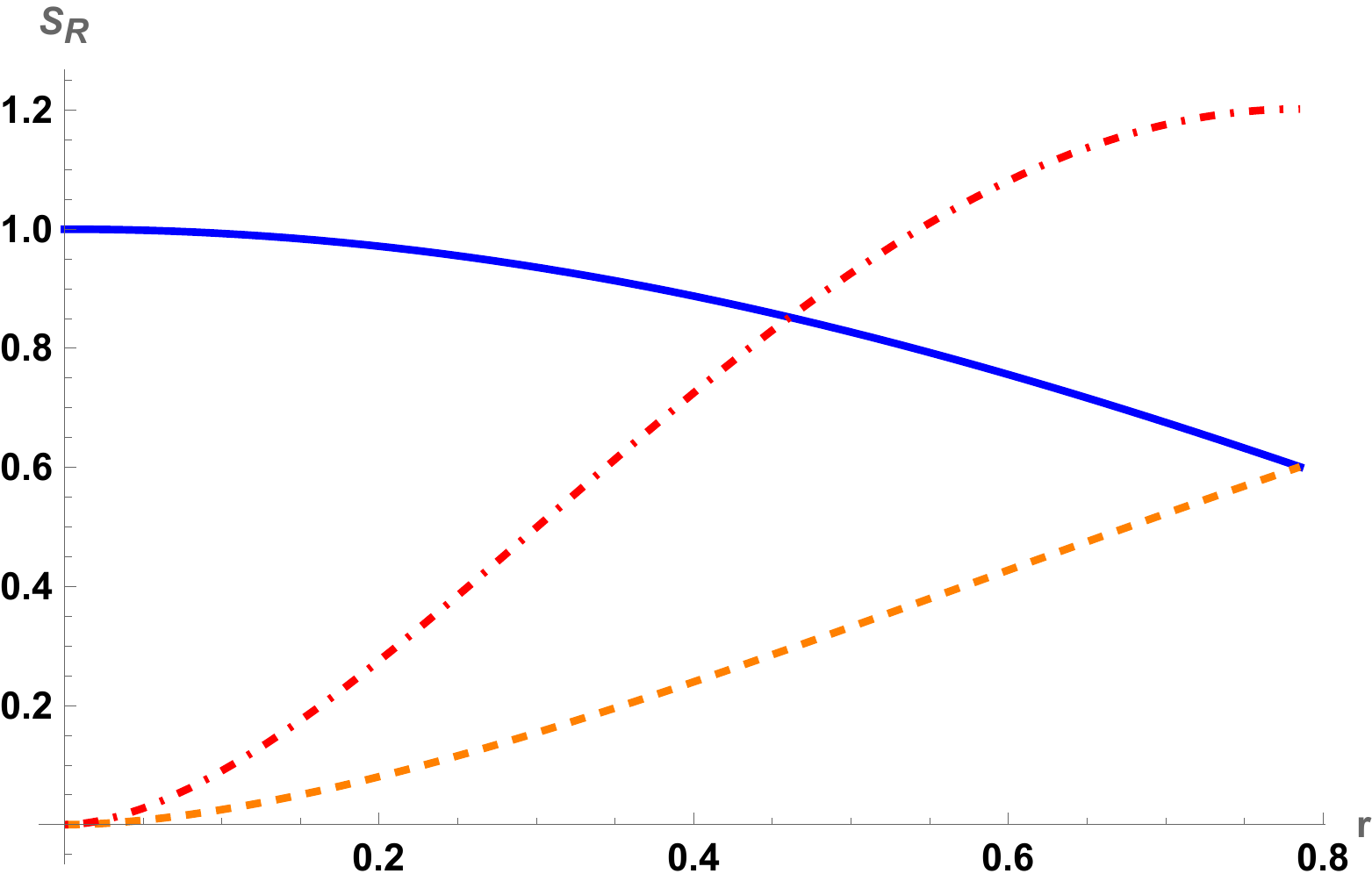}
	\caption{GHZ-state: $S_R(A:B)$ (Blue solid curve),  $S_R(A:\bar{B})$ (orange dashed curve), and  $S_R(B:\bar{B})$ (red dot-dashed curve).  In all  plots that follow for GHZ-state, $\a$ is fixed to its maximally entangled value for the state unless otherwise stated.
	 }
	\label{fig:W_SR_2d}
\end{subfigure}
	\caption{Reflected entropy for Bell, Werner and GHZ state are plotted as a function acceleration.}
\end{figure}

Note that in the inertial frame $r=0$, and $\alpha=\frac{1}{\sqrt{2}}$ correspond to the case of maximally entangled Bell and GHZ state, and  $\alpha=\frac{1}{\sqrt{3}}$ for the maximally entangled W-state. In \cref{fig:Bell_SR_2d,fig:W_SR_2d,fig:GHZ_SR_2d}, we plot $S_R(A:B)$, $S_R(A:\bar{B})$, and  $S_R(B:\bar{B})$ as a function of acceleration $r$ for fixed $\alpha$ for  Bell, W and GHZ state respectively. We notice that $S_R(A:B)$ decreases whereas  $S_R(A:\bar{B})$ increases due to Unruh effect for all the three cases. Furthermore, in the infinite acceleration limit they both reach at the same non-vanishing final value, which indicates that the observers $B$ and $\bar{B}$ becomes indistinguishable at this limit. We notice that as the correlation between $AB$ decreases, the correlation between $A\bar{B}$ grows which is due to the fact of the correlation sharing.   Indeed, this phenomena has also been observed for other entanglement measures as well e.g., entanglement negativity \cite{Alsing:2006cj}. On the other hand $S_R(B:\bar{B})$ increases monotonically starting from zero at $r=0$ culminating to a final non-zero value in the infinite acceleration limit where $r=\frac{\pi}{4}$.

Let us now briefly discuss some of the recent developments which raised a concern  on the generic validity and applicability of the reflected entropy as a correlation measure in quantum information theory.  It has been recently noticed \cite{Hayden:2023yij} that in a qutrit-qutrit-qubit system there exist quantum states which violate the monotonicity of reflected entropy under the operation of partial trace.  Nevertheless,  it remains an important quantity in the context of holography where entanglement wedge cross section is considered as a bulk dual of reflected entropy \cite{Dutta:2019gen}.  Therefore, utilizing the nesting property of the entanglement wedge, it can be argued that reflected entropy in holographic CFT does not suffer from non-monotonicity \cite{Dutta:2019gen,Wall:2012uf}. However, for our states in this work it is essential to confirm that the reflected entropy does reduce under the partial tracing of the degrees of freedom.

\begin{figure}[h]
	\centering
	\begin{subfigure}{.45\textwidth}
		\centering
		\includegraphics[width=.9\linewidth]{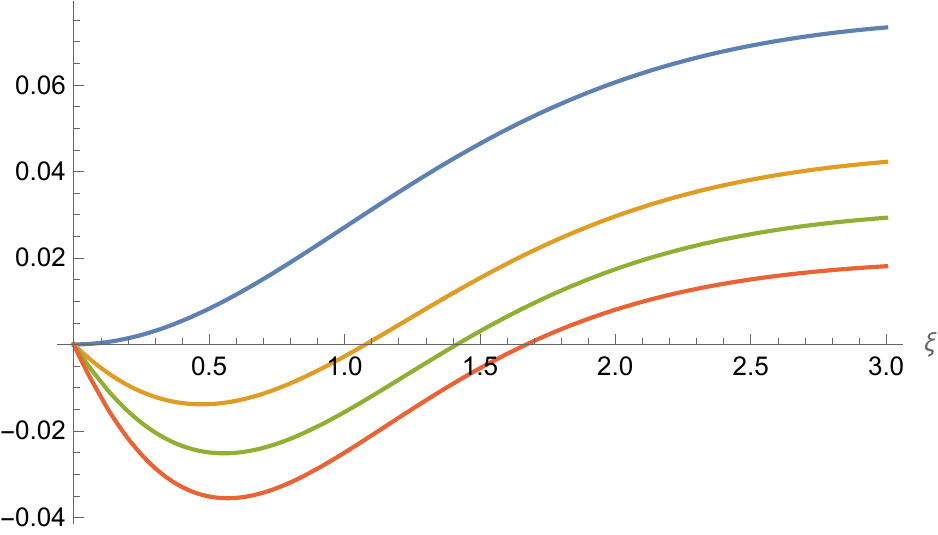}
		\caption{\textbf{$S^{\xi}_{R}(A:BC)-S^{\xi}_{R}(A:B)~vs~\xi$}. Here, $a=.5$ (blue), $a=.75$ (orange), $a=1$ (green), $a=1.5$ (red) and $b=.5$.}
		\label{fig_problem}
	\end{subfigure}\hspace{.5cm}
	\begin{subfigure}{.45\textwidth}
		\centering
		\includegraphics[width=.9\linewidth]{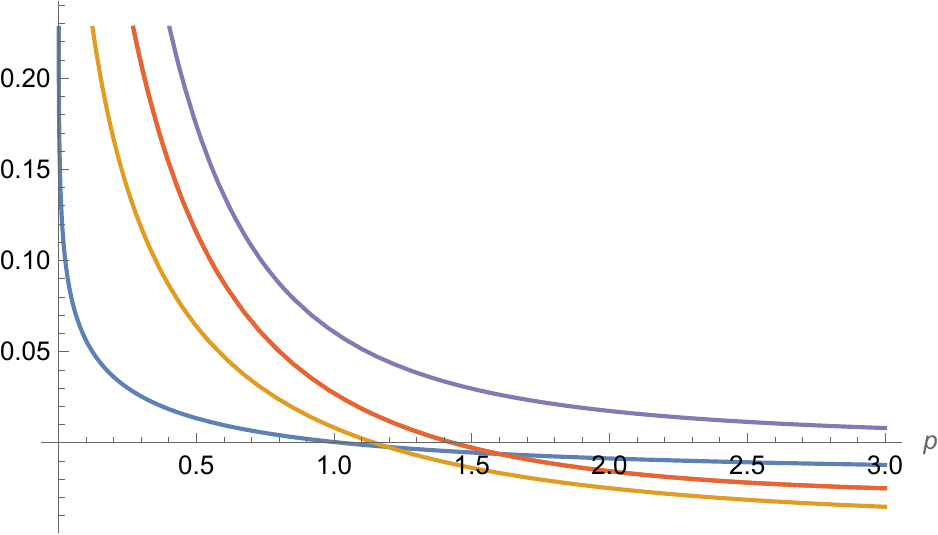}
		\caption{\textbf{$S^{\xi}_{R}(A:BC)-S^{\xi}_{R}(A:B)~vs~p$}. Here, $\xi=.1$ (blue), $a=.5$ (orange), $a=1$ (red), $a=2$ (violet)}
		\label{fig_problem_2}
	\end{subfigure}
	\caption{Monotonicity of  Renyi reflected entropy under partial tracing for the state in \cref{rhoABC}.}
\end{figure}

As a side development to this task we confirm and extend the work of \cite{Hayden:2023yij}, by showing that there exists another  three party state in the Hilbert space $\mathcal{H}_A\otimes\mathcal{H}_B\otimes\mathcal{H}_C=\mathbb{C}^{4}\otimes\mathbb{C}^{3}\otimes\mathbb{C}^2$ which violets the monotonicity of the $\xi$-th Renyi reflected entropy in the domain $\xi\in(0,2)$
\begin{equation}
	\begin{aligned}
		\rho_{A B C}= \frac{1}{6a+2b}\Big[a\Big(
		|000\rangle\langle 000|+| 110\rangle\langle 110|+
		|200\rangle\langle 000|+| 210\rangle\langle 110|+&
		|300\rangle\langle 000|+| 310\rangle\langle 110|\Big)\\&+b\Big(|020\rangle\langle 020|+|121\rangle\langle 121|\Big)\Big].
	\end{aligned}\label{rhoABC}
\end{equation}
In the above expression, $a$ and $b$ are two parameters which can be treated as classical probabilities. Using this state in \cref{rhoABC}, one can compute $\xi$-th Renyi reflected entropy and check the monotonicity under partial trace. It is observed that for some fixed range of parameters $a$ and $b$, the quantity $S^{\xi}_{R}(A:BC)-S^{\xi}_{R}(A:B)$ becomes negative (\cref{fig_problem}). It is easy to check the conditions numerically which yields that $a$ should be larger than $b$. Similar to \cite{Hayden:2023yij}, increasing value of $\frac{a}{b}$ pushes the region of violation towards $\xi=2$. Furthermore, for a fixed value of $\xi$, it can be observed in \cref{fig_problem_2} that the violation of monotonicity occurs at different values of the ratio $p=\frac{a}{b}$. The state \cref{rhoABC} can be generalized for Hilbert space with arbitrary dimensions i.e. $\mathcal{H}_A\otimes\mathcal{H}_B\otimes\mathcal{H}_C=\mathbb{C}^{n+1}\otimes\mathbb{C}^{m+1}\otimes\mathbb{C}^2$ where violation of monotonicity is observed for  Rényi reflected entropy as we show in \cref{monotonicity}.
For the states considered in this study, we are able to confirm that the reflected entropy does reduce under the partial tracing of the degrees of freedom. We include some representative results in \cref{monotonicity}. Consequently, we argue that reflected entropy is a good correlation measure  for our states and the non-inertial observers in  our setup.

\subsection{Bounds of reflected entropy}
\begin{figure}[H]
	\centering
	\begin{subfigure}{.45\textwidth}
		\centering
		\includegraphics[width=.9\linewidth]{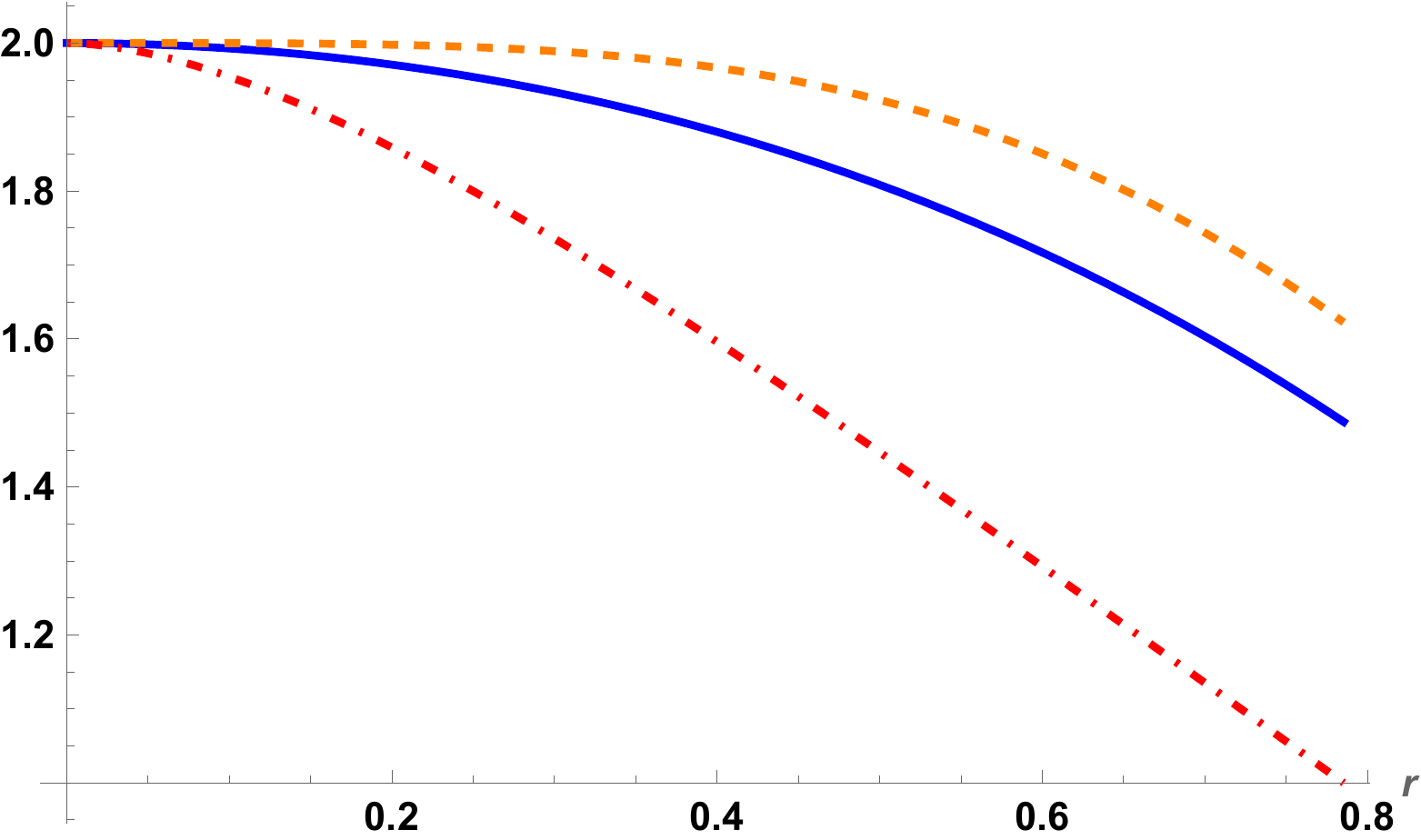}
		\caption{$S_R(A:B)$ (blue solid curve), $min\{2S_A,2S_B\}$ (orange dashed curve), $I(A:B)$ (red dot-dashed).}
		\label{fig:SRMIENWBFvsr}
	\end{subfigure}\hspace{.5cm}
	\begin{subfigure}{.45\textwidth}
		\centering
		\includegraphics[width=.9\linewidth]{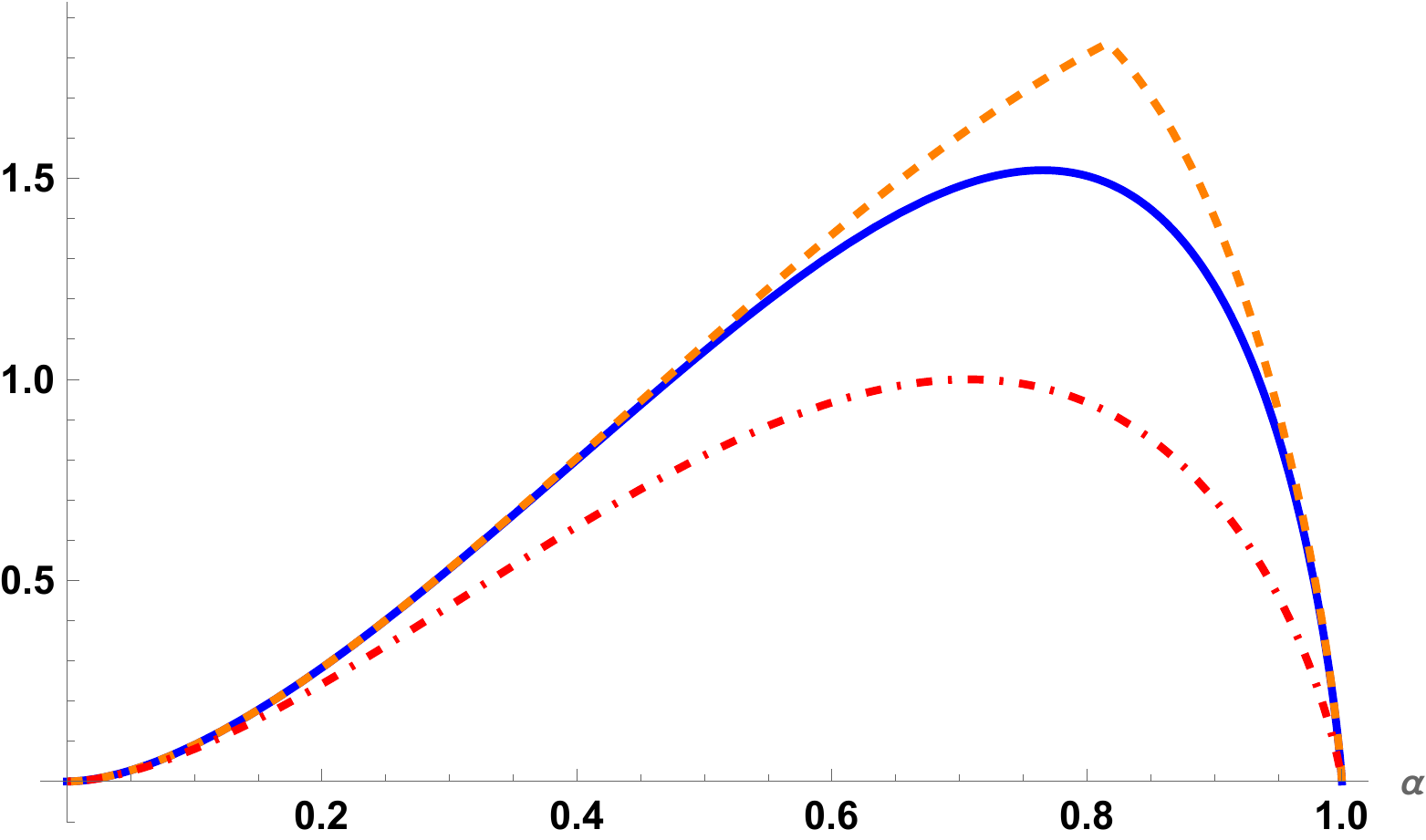}
		\caption{$S_R(A:B)$ (blue solid curve), $min\{2S_A,2S_B\}$ (orange dashed curve), $I(A:B)$ (red dot-dashed).}
		\label{fig:SRMIENWBFvsalpha}
	\end{subfigure}
	\caption{$(a)$  Reflected entropy $S_R(A:B)$ for maximally entangled Bell state as function of $r$ is compared with its upper and lower bound. $(b)$ Reflected entropy $S_R(A:B)$ as a function of $\alpha$ for  $r=\frac{\pi}{4}$ is compared with its upper and lower bound.  }
	\label{fig:Bell_SRMIENWBF}
\end{figure}

\begin{figure}[h]
	\centering
	\begin{subfigure}{.45\textwidth}
		\centering
		\includegraphics[width=.9\linewidth]{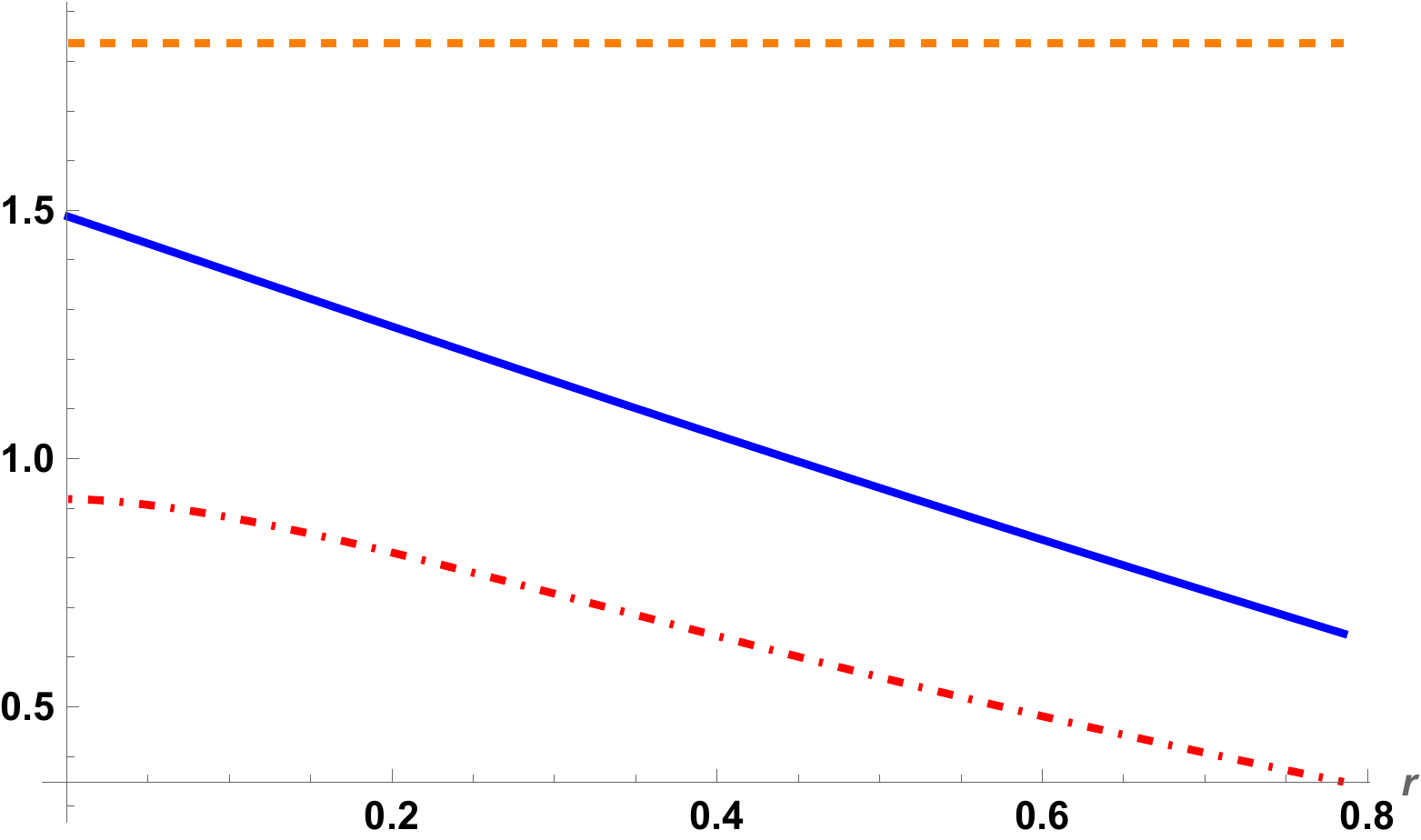}
		\caption{$S_R(A:B)$ (blue solid curve), $min\{2S_A,2S_B\}$ (orange dashed curve), $I(A:B)$ (red dot-dashed curve)}
		\label{fig:W_inequality_vs_r}
	\end{subfigure}\hspace{.3cm}
	\begin{subfigure}{.45\textwidth}
		\centering
		\includegraphics[width=.9\linewidth]{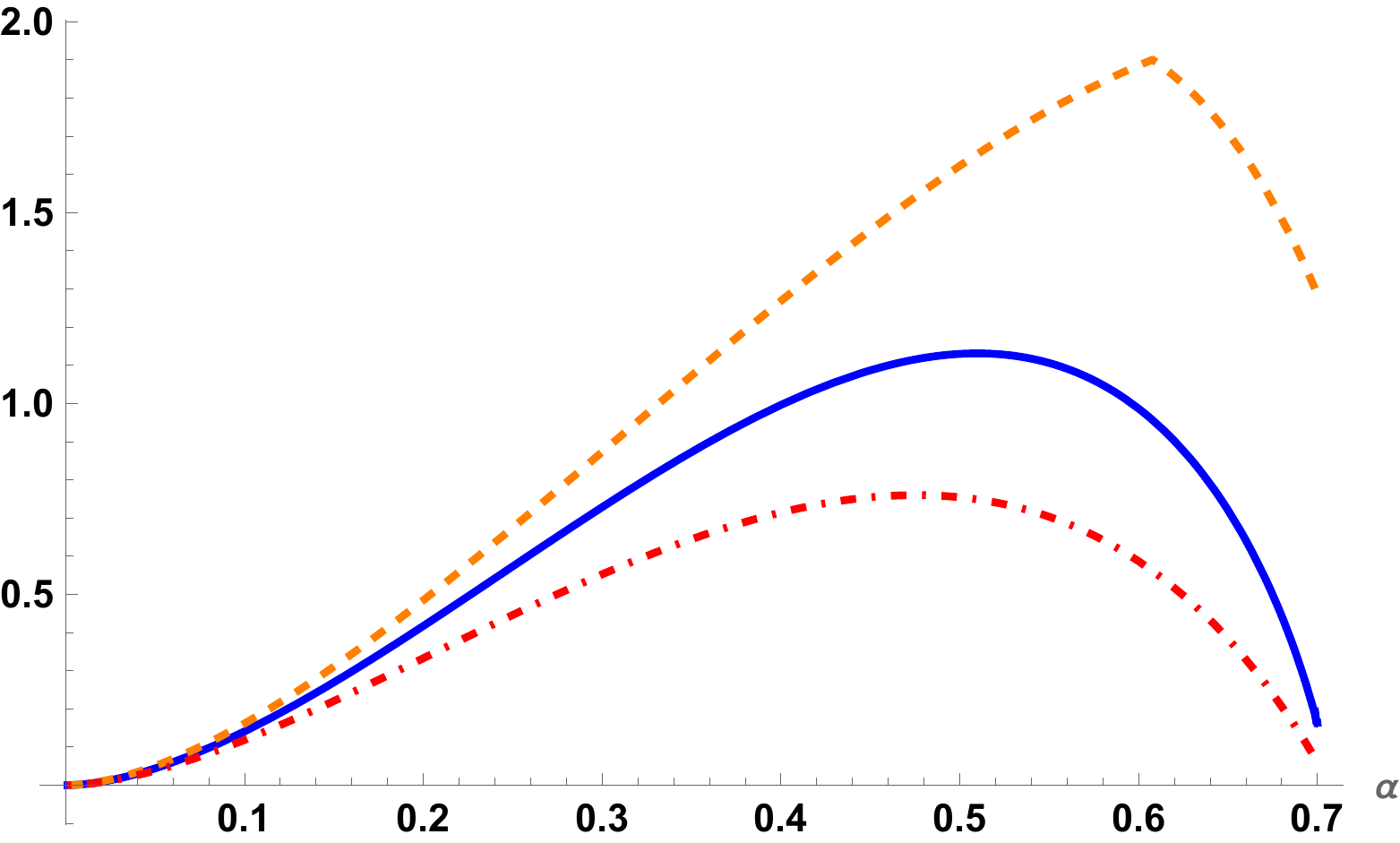}
		\caption{$S_R(A:B)$ (blue solid curve), $min\{2S_A,2S_B\}$ (orange dashed curve), $I(A:B)$ (red dot-dashed curve)}
		\label{fig:W_inequality_vs_alpha}
	\end{subfigure}
	\caption{$(a)$  The reflected entropy $S_R(A:B)$ for the maximally entangled W-state as function of $r$ is compared with its upper and lower bound. $(b)$ Reflected entropy $S_R(A:B)$ as a function of $\alpha$ for  $r=\frac{\pi}{8}$ is compared with its upper and lower bound.  }
	\label{fig:W_SRMIENWBF}
\end{figure}

\begin{figure}[h]
	\centering
	\begin{subfigure}{.45\textwidth}
		\centering
		\includegraphics[width=.9\linewidth]{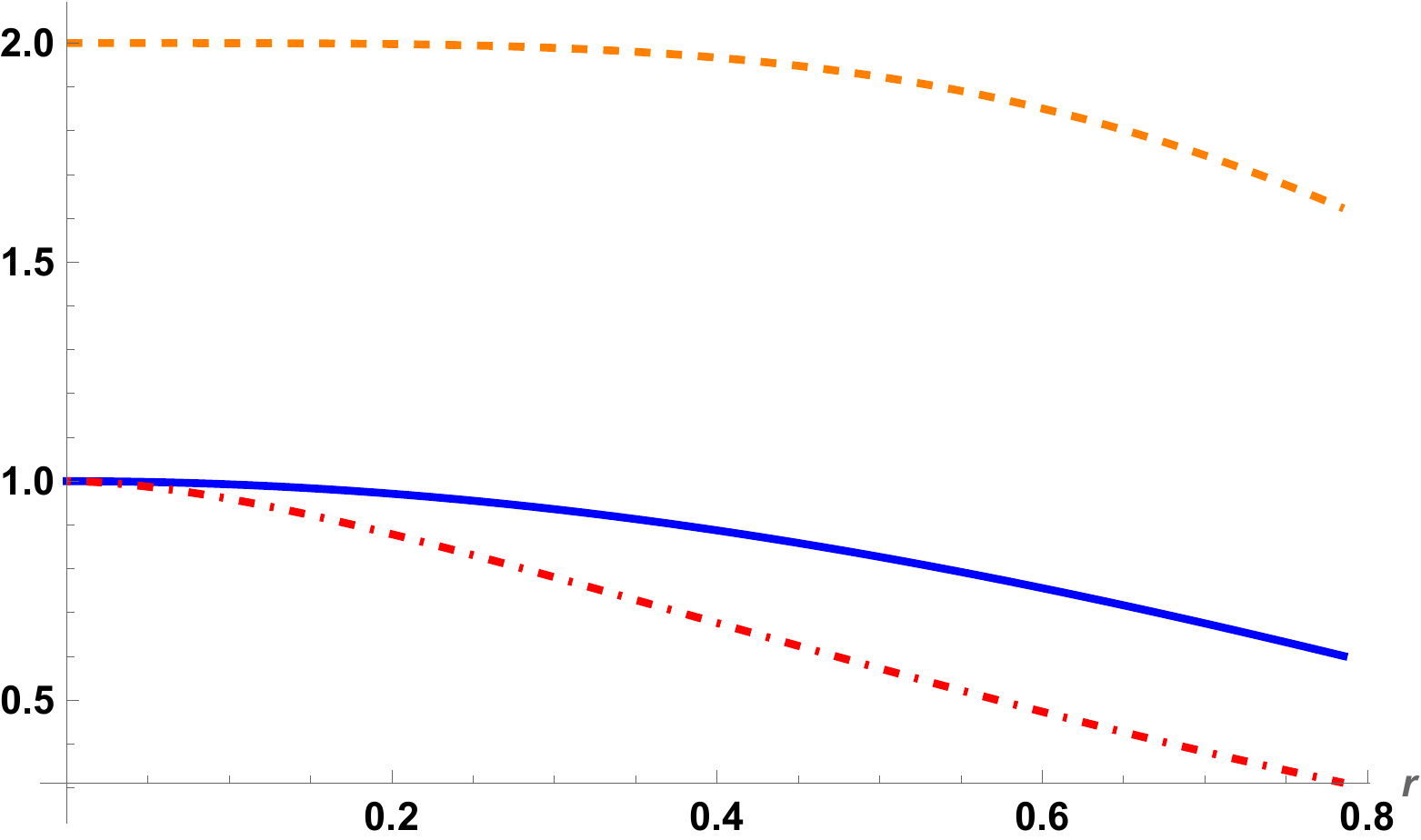}
		\caption{$S_R(A:B)$ (blue solid curve), $min\{2S_A,2S_B\}$ (orange dashed curve), $I(A:B)$ (red dot-dashed curve)}
		\label{fig:GHZ_inequality_vs_r}
	\end{subfigure}\hspace{.3cm}
	\begin{subfigure}{.45\textwidth}
		\centering
		\includegraphics[width=.9\linewidth]{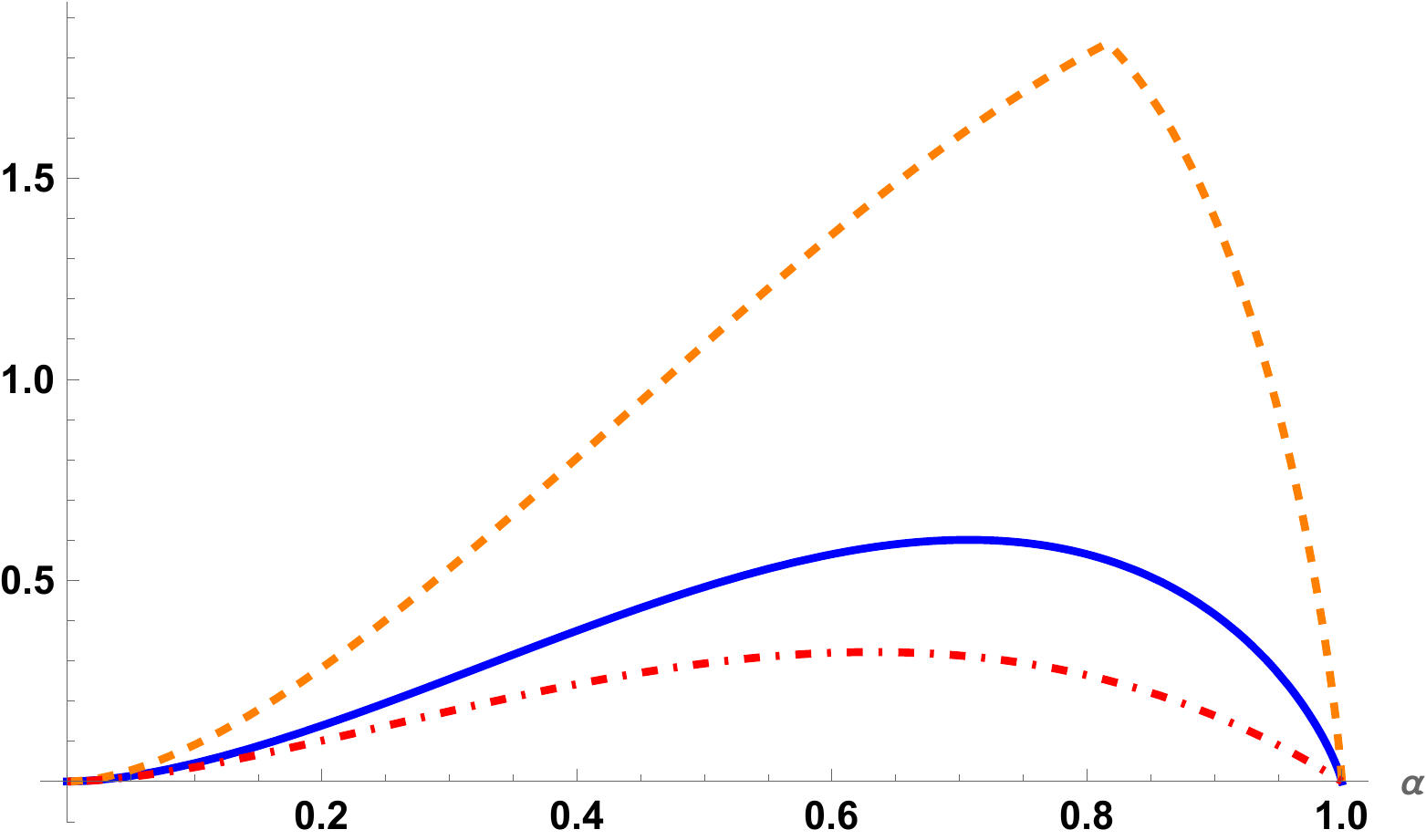}
		\caption{$S_R(A:B)$ (blue solid curve), $min\{2S_A,2S_B\}$ (orange dashed curve), $I(A:B)$ (red dot-dashed curve)}
		\label{fig:GHZ_inequality_vs_alpha}
	\end{subfigure}
	\caption{$(a)$  The reflected entropy $S_R(A:B)$ for the maximally entangled GHZ state as function of $r$  is compared with its upper and lower bound. $(b)$ Reflected entropy $S_R(A:B)$ as a function of $\alpha$ for  $r=\frac{\pi}{4}$ is compared with its upper and lower bound.  }
	\label{fig:GHZ_SRMIENWBF}
\end{figure}

In \cref{fig:Bell_SRMIENWBF,fig:W_SRMIENWBF,fig:GHZ_SRMIENWBF}, we provide an illustrative representation of the upper and lower bound followed by $S_R(A:B)$ as mentioned in \cref{SR_bound} for Bell, Werner and GHZ states respectively. For the case of Bell state, the density matrix $\rho_{AB}$ at $r=0$ is pure and entangled hence reflected entropy $S_R(A:B)$ saturates both upper and lower bounds. Interestingly, increasing $r$ induces tripartite entanglement into the system which leads to the non-saturation of the bound depicted in \cref{fig:SRMIENWBFvsr}. In \cref{fig:SRMIENWBFvsalpha} we observe that for $\alpha=0$ and $\alpha=1$ both the bounds are saturated as expected and near $\alpha=0$, $S_R(A:B)$ (blue solid curve) follows closer the $I(A:B)$ (red dot-dashed curve) whereas close to $\alpha=1$, it follows closer the $min\{2S_A,2S_B\}$ (orange dashed curve). We notice the clear shift of dominance between $min\{2S_A,2S_B\}$ (orange dashed curve) close to $\alpha\simeq .8$ from $2S_A$ to $2S_B$, where the exact value $\alpha$ that this happens depends on the parameters we chose for the state.
On the other hand for the W-state,  $\rho_{AB}$ at $r=0$ is mixed and entangled, as a result none of the bounds are saturated indicating the existence of tripartite entanglement which increases with $r$  \cref{fig:W_inequality_vs_r}. In \cref{fig:W_inequality_vs_alpha} we see that for $\alpha=0$ both the bounds are saturated and at $\alpha=1$ only the lower bound is saturated. We also observe that unlike Bell state, for W-state $S_R(A:B)$ (blue solid curve) near $\alpha=0$ follows close the $min\{2S_A,2S_B\}$ (orange dashed curve)  whereas close to $\alpha=1/\sqrt{2}$, it comes closer to the $I(A:B)$ (red dot-dashed curve).
 Furthermore, we observe a change of dominance in $min\{2S_A,2S_B\}$ from $2 S_A$ to  $2 S_B$ near  $\alpha\simeq.6$, as in the previous case.
As for the GHZ state,   at $r=0$ the density matrix $\rho_{AB}$ is mixed and separable hence only the lower bound is saturated. With increasing $r$ the reflected entropy $S_R(A:B)$ (blue solid curve) decreases and none of the bounds are saturated at large $r$ as it can be seen at \cref{fig:GHZ_inequality_vs_r}. This refers to the existence of the tripartite entanglement at finite $r$. When $S_R(A:B)$ is plotted as a function of $\alpha$ at a fixed $r$, we observe that both the bound are saturated at $\alpha=0$ and $\alpha=1$ presented in \cref{fig:GHZ_inequality_vs_alpha}. Notice the clear change of dominance of $min\{2S_A,2S_B\}$ (orange dashed) near $\alpha\simeq .8$ from $2S_A$ to $2S_B$.

%
\section{Markov gap }\label{MG}

In this section we will study the Markov gap $\emph{h}$ which is proposed as a measure of tripartite entanglement \cite{Zou:2020bly}. For a bipartite system $A\cup B$, it is defined as the difference between  reflected entropy and mutual information \cite{Hayden:2021gno,Liu:2023pdz,Siva:2021cgo}
\begin{equation}\label{Markov_gap}
	h(A: B)=S_R(A: B)-I(A: B)~.
\end{equation}
This quantity is identified with conditional mutual information \cite{Dutta:2019gen}
\begin{equation}\label{hmarkov}
	h(A: B)=I\left(A: B^{\star} \mid B\right)=I\left(B: A^{\star} \mid A\right)~,
\end{equation}
where the conditional mutual information is defined in terms of the linear combination of entanglement entropies as follows
\begin{equation}
	I(A: C \mid B)=S(A B)+S(B C)-S(A B C)-S(B)=I(A: B C)-I(A: B)~.
\end{equation}
The fidelity of a Markov recovery process is related to the conditional mutual information   as \cite{fawzi2015quantum}
\begin{equation}\label{markov}
\max _{\mathcal{R}_{B \rightarrow B C}} F\left(\rho_{A B C}, \mathcal{R}_{B \rightarrow B C}\left(\rho_{A B}\right)\right)\geq \exp^{-I(A: C \mid B)}~.
\end{equation}
Here the Markov recovery process is understood as a technique to obtain the state $\rho_{ABC}$ from any of its bipartite reduced state using Markov recovery map $ \mathcal{R}_{B \rightarrow B C}$\footnote{The Markov recovery map essentially is a quantum channel which produces a bipartite system from a single party system.}. The quantity $F$ in \cref{markov} is known as quantum fidelity which for two density matrices $\rho$ and $\sigma$ is defined as
\begin{equation}
 F(\rho, \sigma)=[\operatorname{Tr} \sqrt{\sqrt{\rho} \sigma \sqrt{\rho}}]^2~.
\end{equation}
Note that it is symmetric in its arguments which lies in the range $0 \leq F(\rho, \sigma) \leq 1$. Utilizing the canonically purified state $\rho_{ABA^{\star}B^{\star}}$, an inequality can be proposed as \cite{Hayden:2021gno}
\begin{equation}
	h(A:B) \geq-\max _{\mathcal{R}_{B \rightarrow B B^{\star}}} \log F\left(\rho_{A B B^{\star}}, \mathcal{R}_{B \rightarrow B B^{\star}}\left(\rho_{A B}\right)\right),
\end{equation}
where \cref{hmarkov,markov} are used to obtain the above equation.

\begin{figure}[H]
	\centering
	\begin{subfigure}{.45\textwidth}
		\centering
		\includegraphics[width=.9\linewidth]{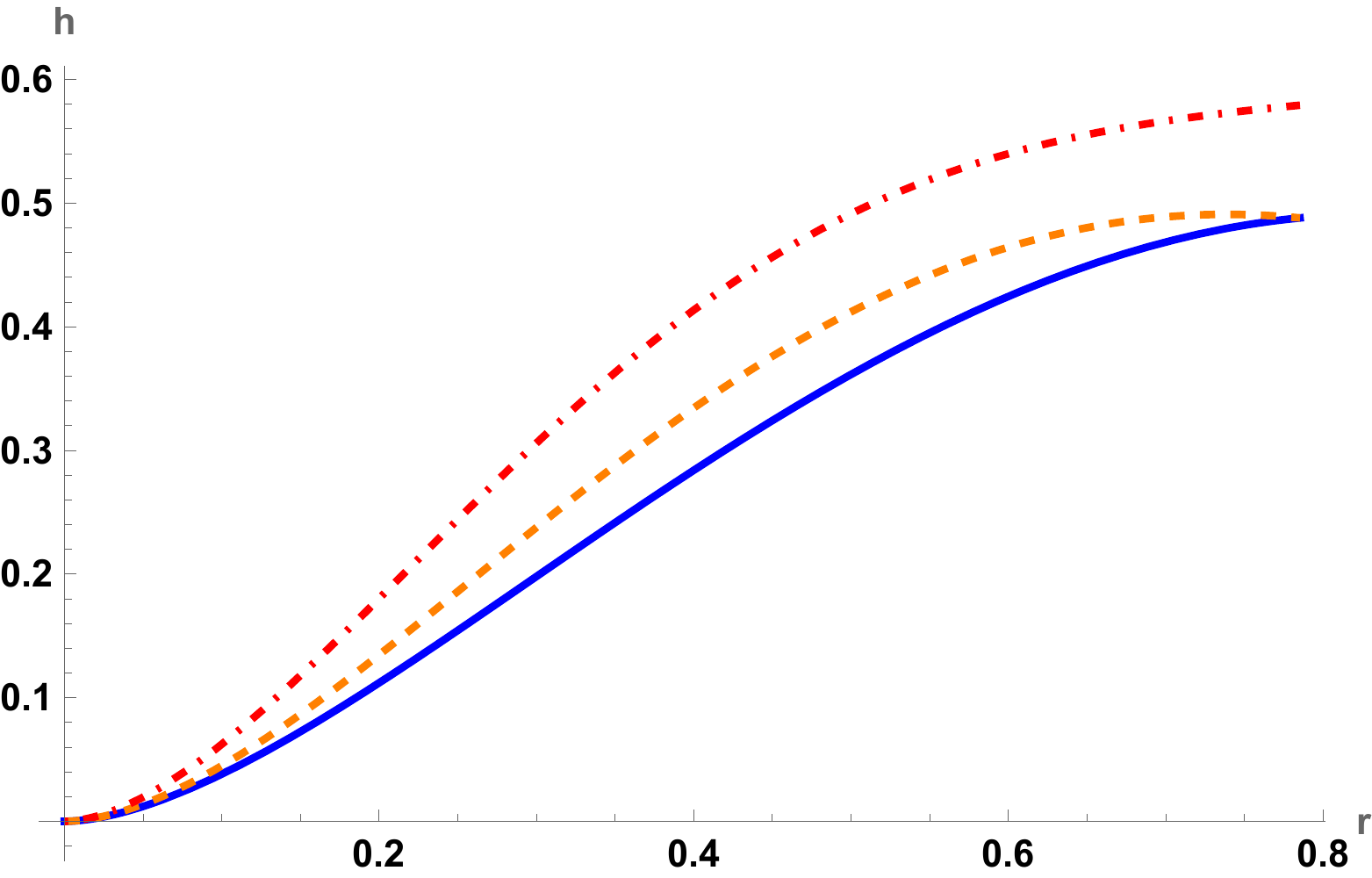}
		\caption{Bell state: $h(A:B)$ (blue solid curve), $h(A:\bar{B})$ (orange dashed curve) and $h(B:\bar{B})$ (red dot-dashed curve).}
		\label{fig:Bell_markovgap_all}
	\end{subfigure}\hspace{1cm}
	\begin{subfigure}{.45\textwidth}
		\centering
		\includegraphics[width=.9\linewidth]{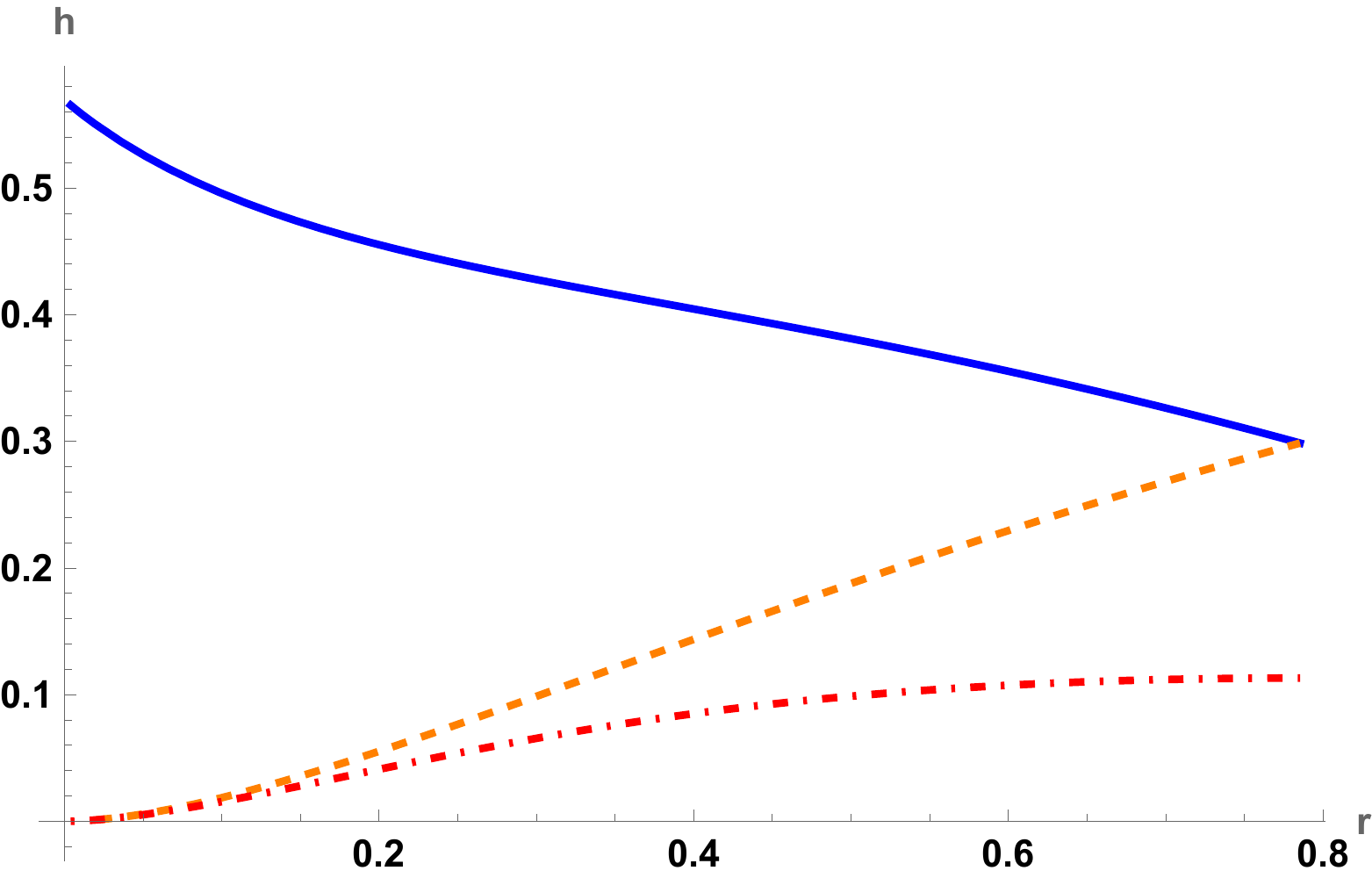}
		\caption{Werner state: $h(A:B)$ (blue solid curve), $h(A:\bar{B})$ (orange dashed curve) and $h(B:\bar{B})$ (red dot-dashed curve).}
		\label{fig:W_markovgap_all}
	\end{subfigure}
\begin{subfigure}{.45\textwidth}
	\centering
	\includegraphics[width=.9\linewidth]{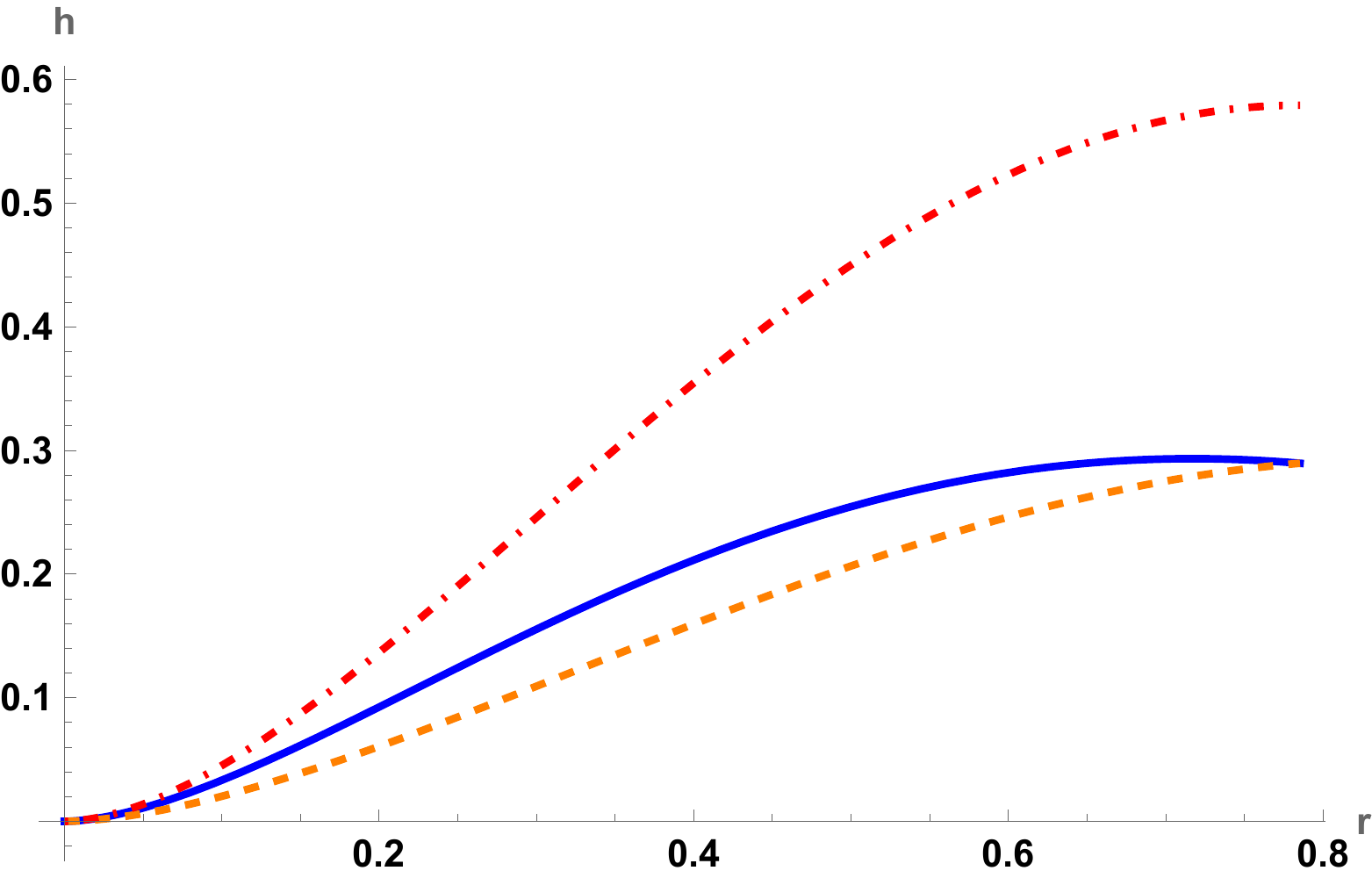}
	\caption{GHZ state: $h(A:B)$ (blue solid curve), $h(A:\bar{B})$ (orange dashed curve) and $h(B:\bar{B})$ (red dot-dashed curve).}
	\label{fig:GHZ_markovgap_all}
\end{subfigure}
	\caption{Markov gap for Bell, Werner and GHZ state are plotted as a function of acceleration. }
	\label{fig:WBFMG}
\end{figure}

Markov gap can be studied in the present setup of this article where we investigate three party (Alice-Bob-Anti Bob) entanglement for Bell, Werner and GHZ states in non inertial frame. The characteristic behavior of the Markov gap $h(A:B)$,  $h(A:\bar{B})$ and $h(B:\bar{B})$ as a function of acceleration $r$ for a constant $\alpha$ are depicted in \cref{fig:WBFMG}. Interestingly, we observe that the Markov gap for all these three cases increase monotonically for Bell state  \cref{fig:Bell_markovgap_all} and GHZ state \cref{fig:GHZ_markovgap_all}, whereas for W-state $h(A:B)$ decreases but  $h(A:\bar{B})$ and $h(B:\bar{B})$ increase monotonically \cref{fig:W_markovgap_all}.

 These figures indicate a few characteristics of multipartite entanglement in these three states. For the Bell state, the entanglement is purely bipartite at $r=0$ and consequently Markov gap vanishes. Anti-Bob $(\bar{B})$ evolves with  increasing acceleration which creates tripartite correlation in the system. As a result, the Markov gaps $h(A:B)$, $h(A:\bar{B})$ and $h(B:\bar{B})$ increases with the acceleration. Interestingly, the authors in \cite{Alsing:2006cj} studied the evolution of three party correlation by exploring a measure named residual tangle\cite{Coffman:1999jd}. Their system under consideration was the same as the first
case in this article i.e. Bell state with accelerating Bob. It was found that the residual tangle is zero for any value of acceleration. This result was interpreted as the absence of tripartite correlation where all the entanglement present in the system is bipartite in nature. As the Markov gap is sensitive towards the the tripartite entanglement, our results can be interpreted as the presence of three party entanglement in the  Bell state under acceleration even if the residual tangle vanishes. This behavior of the Markov gap suggests that it might be able to serve as a fine probe of multipartite entanglement.
Interestingly, on the other hand, W-state has tripartite entanglement between Alice, Bob and anti-Bob in the inertial frame ($r=0$) which is indicated by the non-zero initial value of $h(A:B)$. Furthermore, anti-Bob does not exist in the inertial frame where the Markov gap related to him is zero. The Markov gap $h(A:B)$ shows a monotonic decreasing behavior because of the entanglement sharing between Alice, Bob and anti-Bob  with increasing acceleration. Note that, at $r=\frac{\pi}{4}$, the Markov gap $h(A:B)$ coincides with $h(A:\bar{B})$ similar to other findings in this article and in \cite{Alsing:2006cj}. Furthermore, for GHZ states $h(A:B)$,  $h(A:\bar{B})$ and $h(B:\bar{B})$ increase monotonically as function of $r$ starting from zero at $r=0$. The nature of the tripartite entanglement computed by the Markov gap for GHZ  state are similar to that of Bell state with accelerating Bob as depicted in \cref{fig:GHZ_markovgap_all}.

\section{A monotonic function for reflected entropy}\label{c_function}

In this section we will study few properties of reflected entropy in our setup by defining a specific function of temperature and frequency. 
Here we use the relation between the acceleration and the Unruh temperature,
\begin{equation}
	r=\tan^{-1} (e^{-\frac{\omega}{2T}})~,
\end{equation}
to obtain the characteristics of reflected entropy with respect to the Unruh temperature $T$. We find that for fixed $\omega$ and increasing $T$ all the maximally entangled states have a monotonically decreasing behavior of $S_R(A:B)$ and  $I(A:B)$  with $T$. 
We also notice that the dimensionless, single parameter function $\sigma(T)$ which we define as,
\begin{equation}\label{abc}
	\sigma(T)=\frac{1}{\omega}\frac{\partial S_R}{ \partial \(\frac{1}{T}\)}~,
\end{equation}
where $\omega$ can be considered as the fixed scale, has a monotonic properties with respect to the increase of temperature.
\begin{figure}[h]
	\centering
	\begin{subfigure}{.45\textwidth}
		\centering
		\includegraphics[width=.9\linewidth]{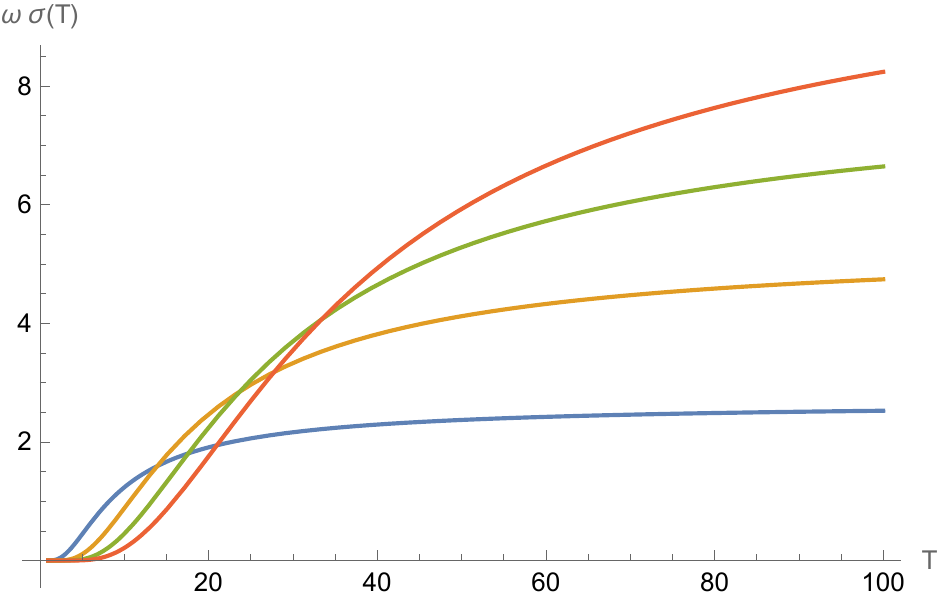}
		\caption{  Bell state: $\omega=10$ (blue), $\omega=20$ (orange), $\omega=30$  (green) and  $\omega=40$  (red). }
		\label{c_bell}
	\end{subfigure}\hspace{0.5cm}%
	\begin{subfigure}{.45\textwidth}
		\centering
		\includegraphics[width=.9\linewidth]{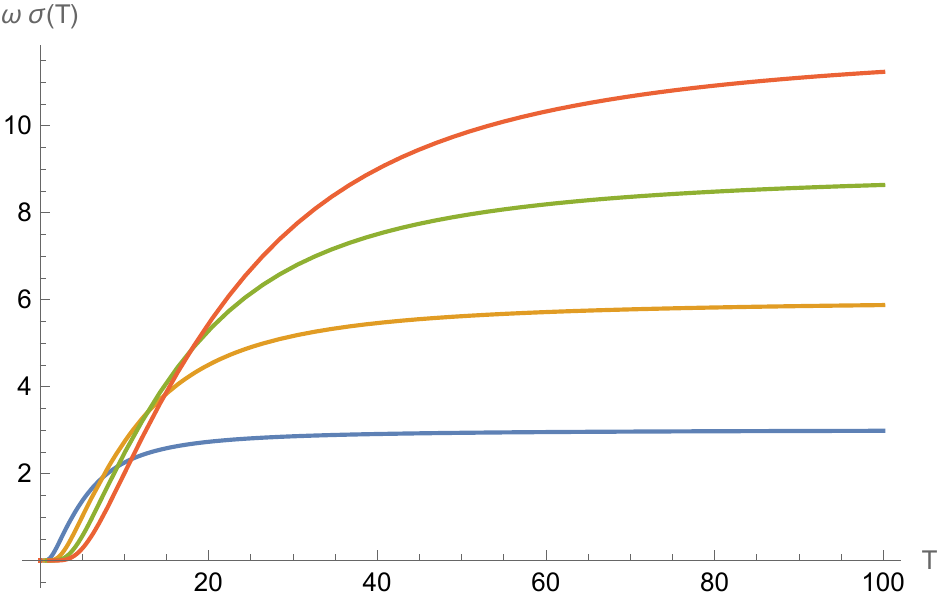}
		\caption{Werner state: $\omega=10$ (blue), $\omega=20$ (orange), $\omega=30$  (green) and  $\omega=40$  (red).}
		\label{c_werner}
	\end{subfigure}
	\begin{subfigure}{.45\textwidth}
		\centering
		\includegraphics[width=.9\linewidth]{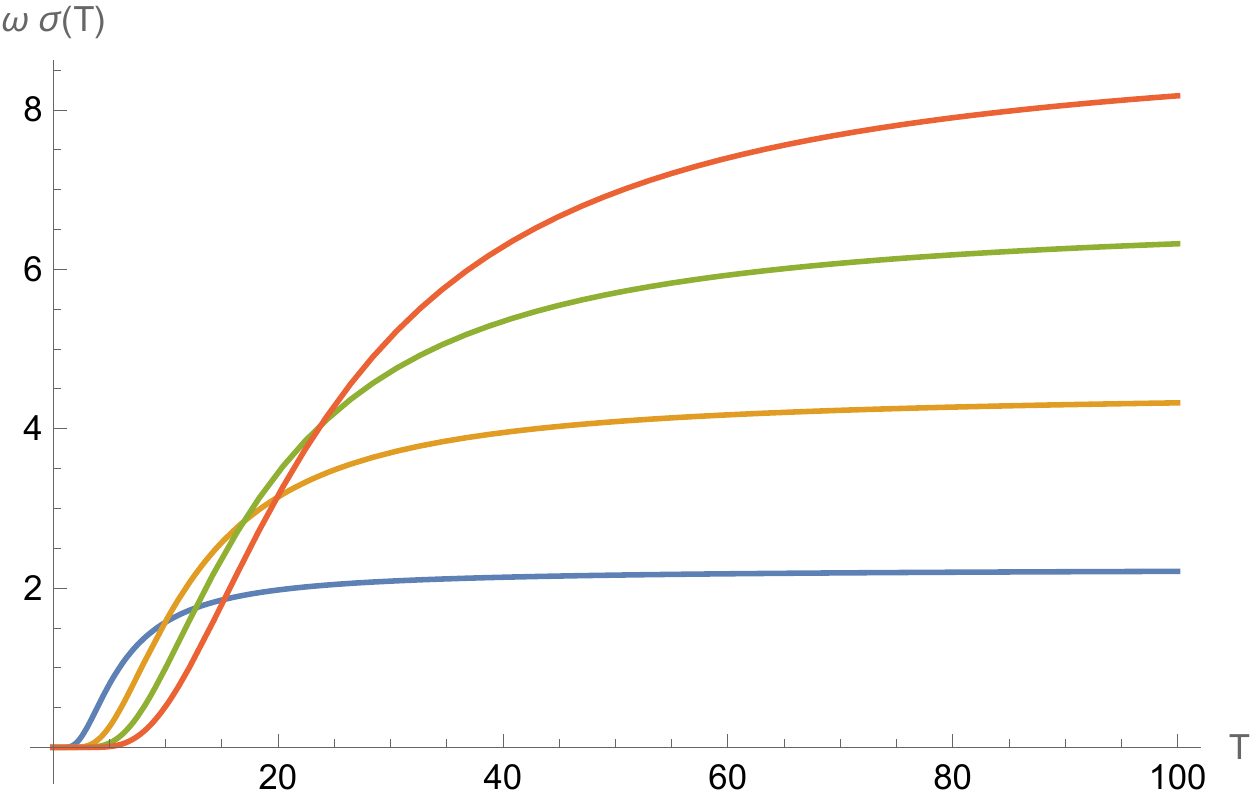}
		\caption{ GHZ state: $\omega=10$ (blue), $\omega=20$ (orange), $\omega=30$  (green) and  $\omega=40$  (red).  }
		\label{ghz-sigma}
	\end{subfigure}
	\caption{ $\omega\cdot \sigma(T)$ as function of Unruh temperature $T$.}
	\label{}
\end{figure}
 In \cref{c_bell,c_werner,ghz-sigma}, we observe $\sigma(T)$ increases monotonically with increasing Unruh temperature, meaning that the entanglement measure $S_R$ we are interested in decreases for increasing acceleration. The $\sigma$-function tends to zero for $T\to 0$ and for $T\to\infty$ saturates to fixed values which are different for each state and independent of $\omega$. 
Notice that the $\sigma$-function does not suffer from divergences and for our states, for two party and three party system having bipartite, W and GHZ type entanglement, the generic behavior remains the same. We point out that the definition of this function is partly motivated by the well known c-function in terms of the entanglement entropy proposed and further studied in \cite{Casini:2004bw,Casini:2006es,Ryu:2006ef,Myers:2012ed,Giataganas:2019uoh}. By establishing the monotonicity of the function, one may like to question whether there exist a clear physically relevant interpretation of the function in relation to the degrees of freedom shared between the two parties. An initial approach is that observers with higher accelerations are further away from the origin,  covering only a subspace of the observers with lower accelerations and therefore should be associated with less degrees of freedom.  It is worthy to be studied in even more complex setups, in order to obtain a more solid interpretation.


\section{Summary and discussion } \label{sum_dis}

In this paper we have investigated the behavior of reflected entropy between two modes of free fermionic fields in a non-inertial frame from the perspective of two relatively accelerated observers. Alice and Bob, for bipartite system described by Bell state, and added Charlie for tripartite system represented by Werner and GHZ state. We confirm that for our 3-qubit and 4-qubit states, Renyi reflected entropy is monotonic under partial trace, allowing us to use reflected entropy as a legitimate measure of correlation. This is an essential check since recent developments raised concerns about the generic validity and applicability of the reflected entropy as a correlation measure in quantum information theory \cite{Hayden:2023yij}, by pointing out the existence of a fine-tuned state that violates the desirable monotonicity. In fact we validate these developments by showing that such fine tuned states can exist in higher dimensional Hilbert spaces and we explicitly present a class of such states.  Nevertheless, getting back to our setup and our used states in this work we confirm that the reflected entropy does reduce under the partial tracing of the degrees of freedom.

We show that the reflected entropy between Alice and Bob degrades with acceleration due to the Unruh effect, culminating in a non-vanishing minimum value. We also computed the Reflected entropy between Alice and anti-Bob (who is causally separated from the observer Bob in region I) and Bob and anti-Bob. We discovered that the reflected entropy increases monotonically with acceleration in these two circumstances.   
Furthermore, we explored the Markov gap, which is a measure of tripartite entanglement, between all three parties Alice-Bob, Alice-anti-Bob, and Bob-anti-Bob. We find that the Markov gap increases monotonically with acceleration in all three scenarios for Bell and GHZ state whereas for W-state  it declines for Alice-Bob but grows for  Alice-anti-Bob, and Bob-anti-Bob. In Bell and GHZ state, for vanishing acceleration, the Markov gap was zero. We have argued that acceleration causes tripartite entanglement in the system for all the three states in consideration, as evidenced by the non-zero value of the Markov gap at finite and even infinite acceleration in \cref{fig:Bell_markovgap_all,fig:W_markovgap_all,fig:GHZ_markovgap_all}. This observation suggests that the Markov gap could be used to characterize the there-body correlation encoded for tripartite states apart from some other measures in the literature.

We have suggested a dimensionless $\sigma$-function of reflected entropy for a fixed mode frequency which  preserves  monotonicity with increasing temperature. Due to the character of the reflected entropy, this specific function is free from any divergences. The function exhibits always a convergence to certain values at $T\to 0$ and $T\to\infty$. We suggest the possibility that this function contains information of the effective degrees of freedom or the shared correlation between two parties.

As for future direction, it would be interesting to ask what happens if Alice and Bob both accelerate simultaneously with different rate of acceleration. Intuitively, one could expect that reflected entropy between Alice and Bob to further decrease, eventually reaching a non-zero value in the infinite acceleration limit. Another interesting path for future research along this line is to address the same question for black hole spacetimes. Besides, it will be exciting to check the generalized properties of the $\sigma$-function independent of the choices of states.

\section{Acknowledgment}
The authors are grateful to V. Malvimat for useful discussions. D.G. would like to thank the Department of Theoretical Physics of CERN for hospitality during the final
stages of this work. The research work of JKB  supported by the National Science and Technology Council of Taiwan with the grant 112-2636-M-110-006. The research work of D.G. is supported by the National Science and Technology Council (NSTC) of Taiwan with the Young Scholar Columbus Fellowship grant 112-2636-M-110-006. This research work of SM and WW are supported in part by the Taiwan’s Ministry of Science and Technology (109-2112-M-033-005-MY3) and the National Center for Theoretical Sciences (NCTS).

\begin{appendices}

\section{The Density Matrices of Bell state } \label{Bell_state}
The density matrices  $\rho_{AB}^{(B)}$, $\rho_{A\bar{B}}^{(B)}$, and  $\rho_{B\bar{B}}^{(B)}$ for the Bell state have been given in the \cref{density_matrix_bell}. Using a proper basis $\{|00\rangle,|01\rangle,|10\rangle,|11\rangle\}$,  $\rho_{AA^{\star}}=\mathrm{Tr}_{BB^{\star}}(|\sqrt{\rho_{ABA^{\star}B^{\star}}}\rangle\langle\sqrt{\rho_{ABA^{\star}B^{\star}}}|)$, $\bar{\rho}_{AA^{\star}}=\mathrm{Tr}_{B\bar{B^{\star}}}(|\sqrt{\rho_{A\bar{B}A^{\star}\bar{B^{\star}}}}\rangle\langle\sqrt{\rho_{A\bar{B}A^{\star}\bar{B^{\star}}}}|)$ and $\rho_{BB^{\star}}=\mathrm{Tr}_{\bar{B}\bar{B^{\star}}}(|\sqrt{\rho_{B\bar{B}B^{\star}\bar{B^{\star}}}}\rangle\langle\sqrt{\rho_{B\bar{B}B^{\star}\bar{B^{\star}}}}|)$ are given as follows
	\begin{equation}\label{Bell_stat_purified_AAstar}
		\left(
		\begin{array}{cccc}
			\frac{\alpha ^2 \left(\left(2 \alpha ^2-1\right) \cos 2r+1\right)}{-\alpha
				^2+\alpha ^2 \cos 2r+2} & 0 & 0 & -\frac{\sqrt{2} \alpha ^2 \left(\alpha
				^2-1\right) \sin ^2 r}{\sqrt{\alpha ^2 \sin ^2r \left(-\alpha ^2+\alpha
					^2 \cos 2r+2\right)}} \\
			0 & -\frac{2 \alpha ^2 \left(\alpha ^2-1\right) \cos ^2r}{-\alpha ^2+\alpha
				^2 \cos 2r+2} & 0 & 0 \\
			0 & 0 & -\frac{2 \alpha ^2 \left(\alpha ^2-1\right) \cos ^2r}{-\alpha
				^2+\alpha ^2 \cos 2r+2} & 0 \\
			-\frac{\sqrt{2} \alpha ^2 \left(\alpha ^2-1\right) \sin ^2r}{\sqrt{\alpha
					^2 \sin ^2r \left(-\alpha ^2+\alpha ^2 \cos 2r+2\right)}} & 0 & 0 &
			\frac{2 \left(\alpha ^2-1\right)^2}{-\alpha ^2+\alpha ^2 \cos 2r+2} \\
		\end{array}
		\right),
	\end{equation}
	\begin{equation}\label{Bell_stat_purified_AAstar}
		\left(
		\begin{array}{cccc}
			\frac{\alpha ^2 \left(\left(2 \alpha ^2-1\right) \cos (2 r)-1\right)}{\alpha ^2+\alpha ^2
				\cos (2 r)-2} & 0 & 0 & -\frac{\alpha  \left(\alpha ^2-1\right) \cos
				(r)}{\sqrt{1-\alpha ^2 \cos ^2(r)}} \\
			0 & \frac{2 \alpha ^2 \left(\alpha ^2-1\right) \sin ^2(r)}{\alpha ^2+\alpha ^2 \cos (2
				r)-2} & 0 & 0 \\
			0 & 0 & \frac{2 \alpha ^2 \left(\alpha ^2-1\right) \sin ^2(r)}{\alpha ^2+\alpha ^2 \cos
				(2 r)-2} & 0 \\
			-\frac{\alpha  \left(\alpha ^2-1\right) \cos (r)}{\sqrt{1-\alpha ^2 \cos ^2(r)}} & 0 & 0
			& -\frac{2 \left(\alpha ^2-1\right)^2}{\alpha ^2+\alpha ^2 \cos (2 r)-2} \\
		\end{array}
		\right),
	\end{equation}
	\begin{equation}\label{Bell_stat_purified_AAstar_appendix}
		\left(
		\begin{array}{cccc}
			\alpha ^2 \cos ^4(r) & 0 & 0 & \alpha  \sqrt{1-\alpha ^2} \cos ^2(r) \\
			0 & \alpha ^2 \sin ^2(r) \cos ^2(r) & 0 & 0 \\
			0 & 0 & \alpha ^2 \sin ^2(r) \cos ^2(r) & 0 \\
			\alpha  \sqrt{1-\alpha ^2} \cos ^2(r) & 0 & 0 & -\alpha ^2+\alpha ^2 \sin ^2(r) \cos
			^2(r)+1 \\
		\end{array}
		\right).
	\end{equation}
	The Reflected entropy $S_R(A:B)$, $S_R(A:\bar{B})$, and  $S_R(B:\bar{B})$   may be obtained by employing the eq. \eqref{reflected_entropy} and using the information above. The expression of these density matrices $\rho_{AA^*}$, $\bar{\rho}_{AA^*}$ and $\rho_{BB^*}$   for W-state and GHZ state are large and we have not included them here for presentation reasons.	

\section{Polygamy inequality }

\begin{figure}[H]
	\centering
	\begin{subfigure}{.3\textwidth}
		\centering
		\includegraphics[width=.82\linewidth]{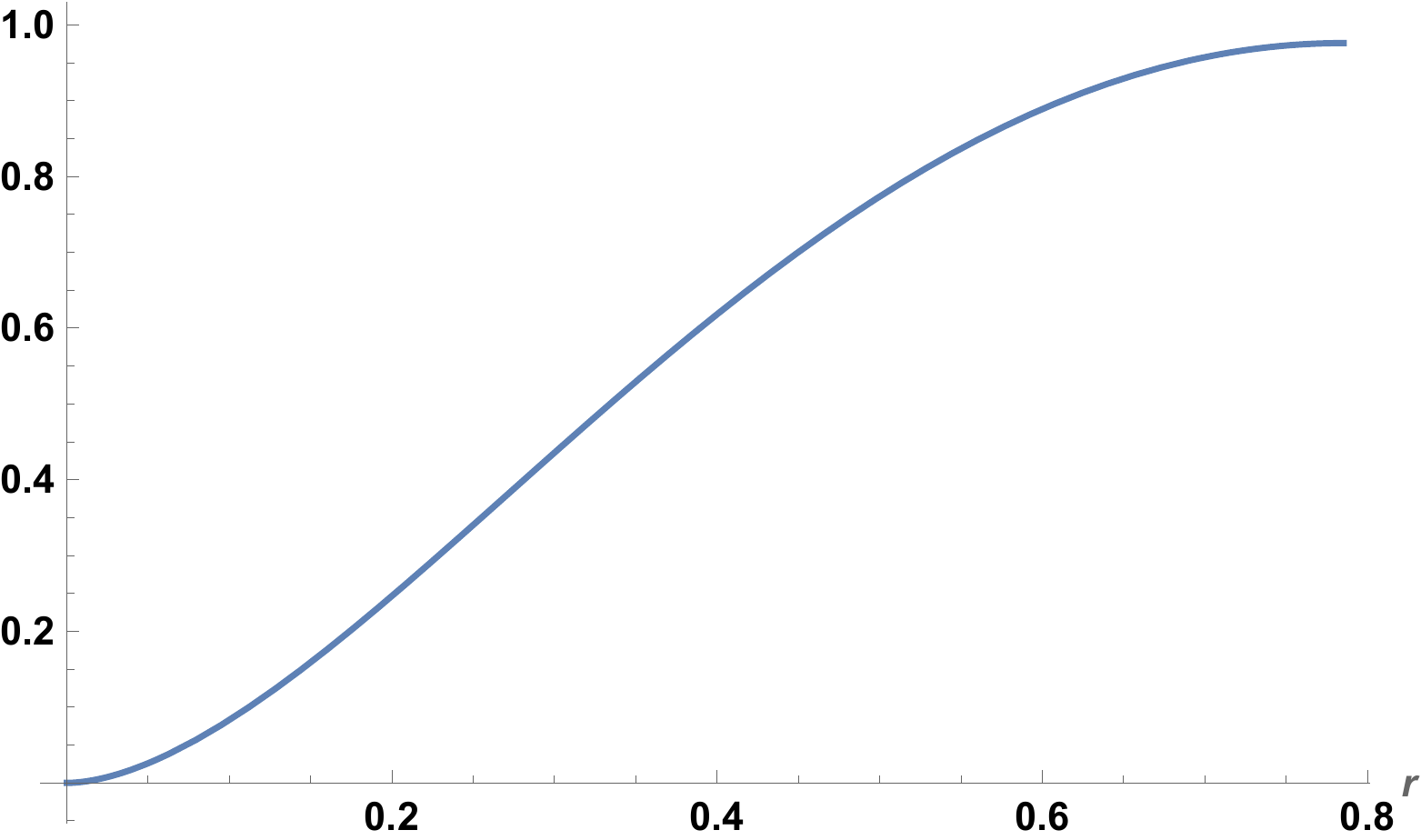}
		\caption{Bell state.
		}
		\label{fig:Bell_poly}
	\end{subfigure}\hspace{.05cm}%
	\begin{subfigure}{.3\textwidth}
		\centering
		\includegraphics[width=.82\linewidth]{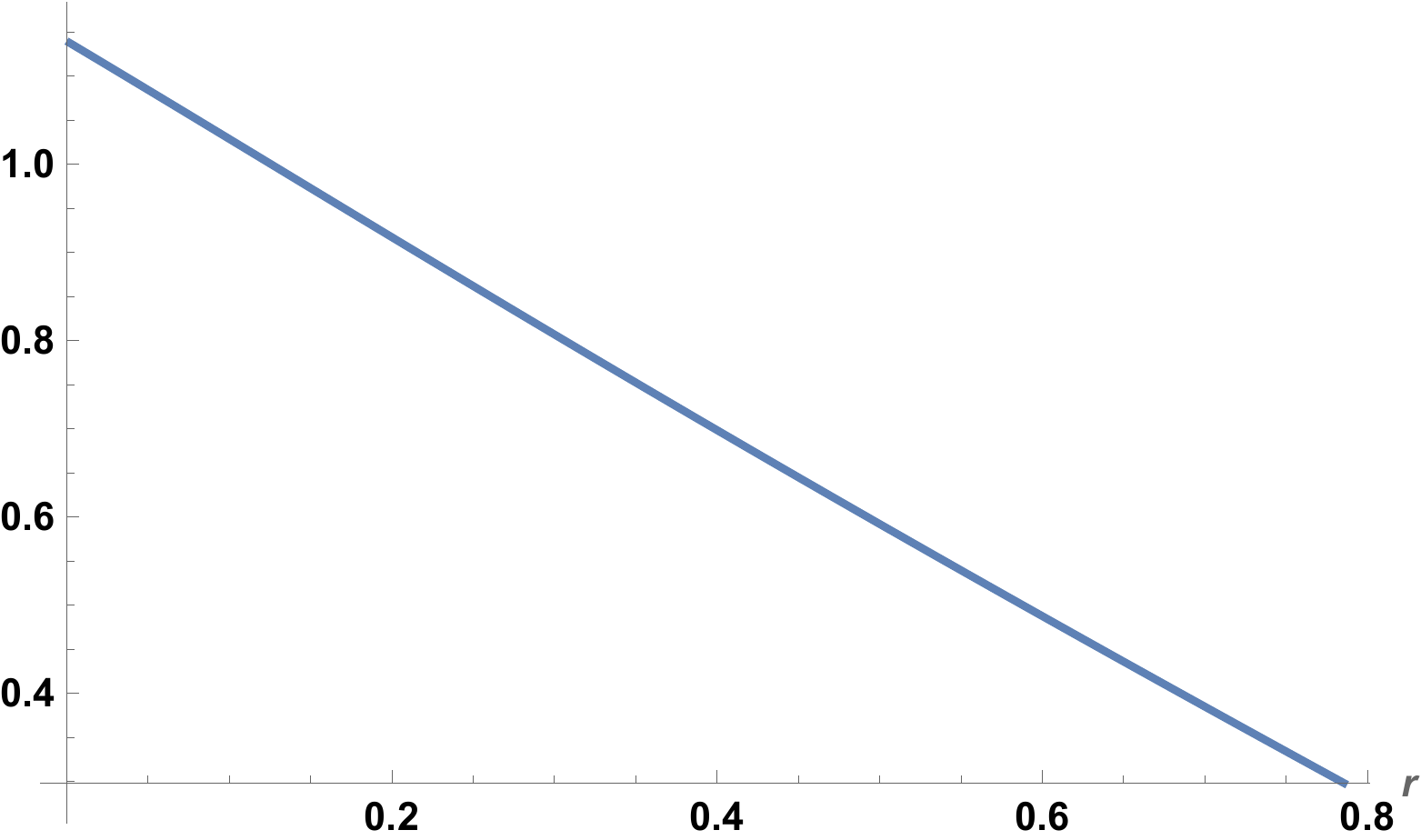}
		\caption{W-state.
		}
		\label{fig:W_poly}
	\end{subfigure}\hspace{.05cm}
\begin{subfigure}{.3\textwidth}
	\centering
	\includegraphics[width=.82\linewidth]{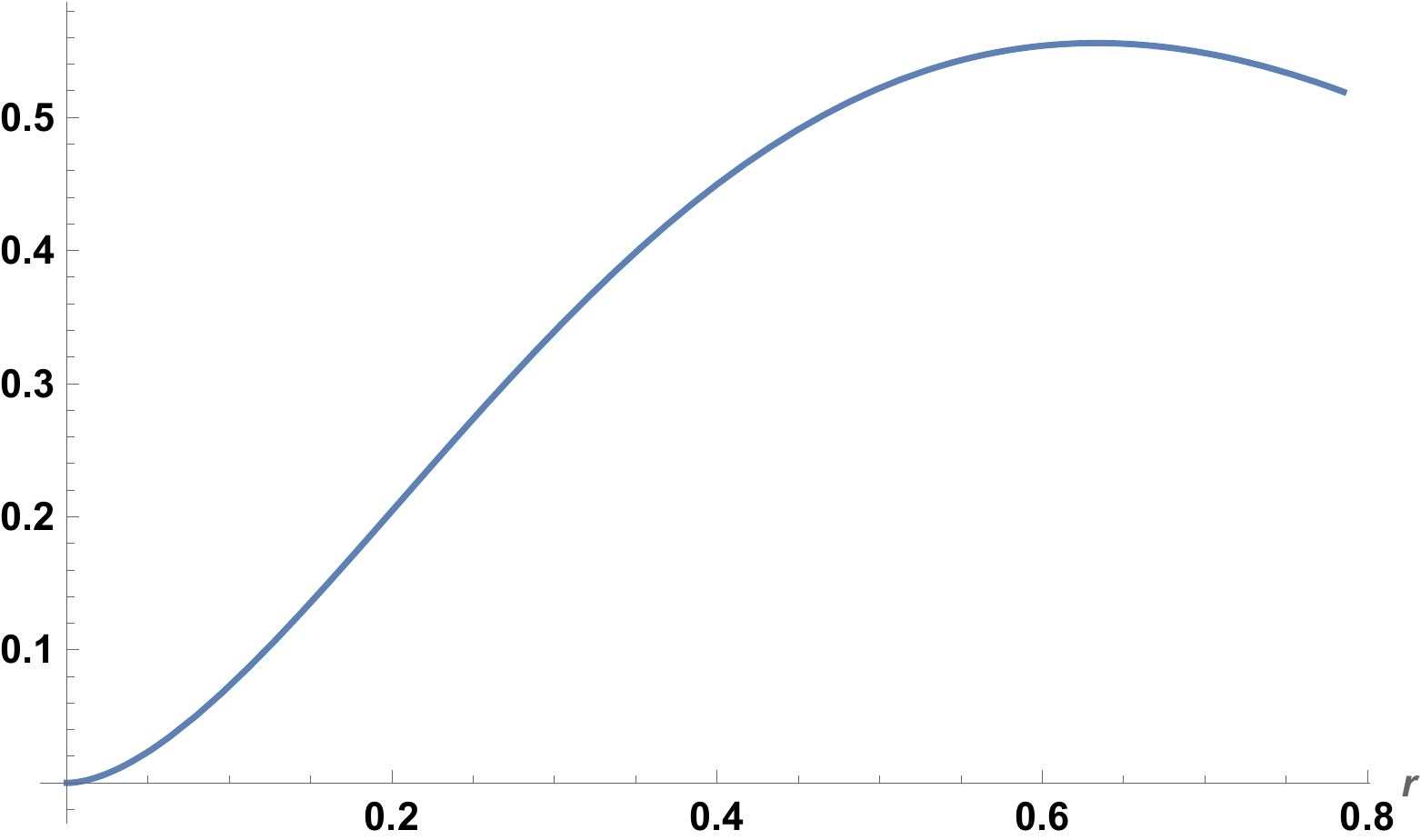}
	\caption{GHZ-state.
	}
	\label{fig:GHZ_poly}
\end{subfigure}
	\caption{ Polygamy inequality as function of acceleration. }
	\label{fig:polygamy_inequality}
\end{figure}

To show the polygamy inequality \cref{polygamy_inequality}, we construct $S_R(A:{B})+S_R(A:\bar{B})-S_R(A:B\bar{B})$ for Bell state $S_R(A:{B})+S_R(A:\bar{B}C)-S_R(A:B\bar{B}C)$ for Werner and GHZ states for fixed $\alpha$ and plot these in \cref{fig:Bell_poly,fig:W_poly,fig:GHZ_poly}. We notice that for Bell and GHZ states it increases monotonically with growing $r$ and remain positive for all value of $r$, thus satisfies the polygamy inequality. Unlike Bell and GHZ states, for W-state it decreases monotonically with $r$ from a maximum value at $r=0$ although satisfies the polygamy inequality as it remains positive for all $r$.

\section{Monotonicity of Reflected Entropy}	\label{monotonicity}
\begin{figure}[H]
	\centering
	\begin{subfigure}{.33\textwidth}
		\centering
		\includegraphics[width=.86\linewidth]{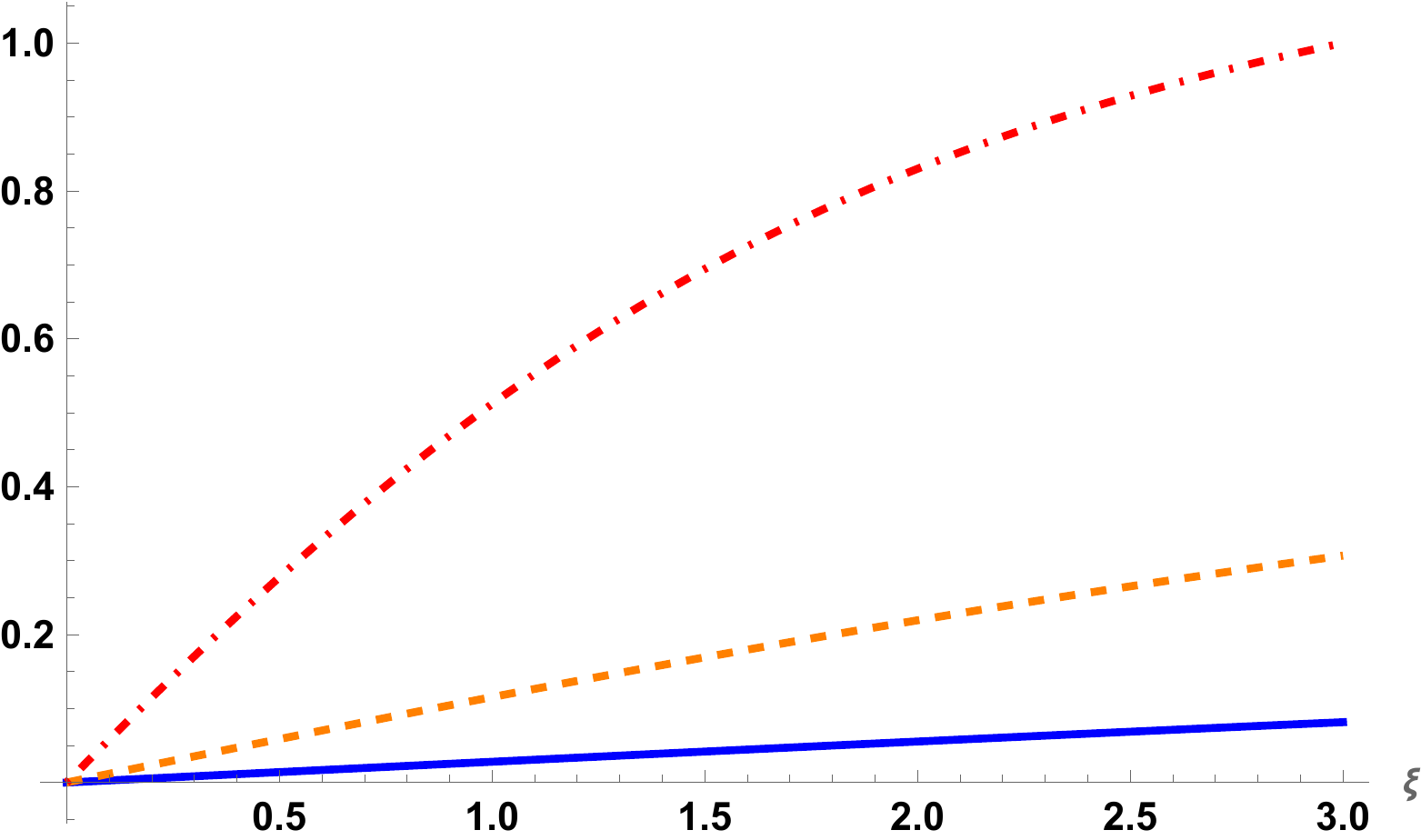}
		\caption{Bell state.
		}
		\label{fig:Bell_mono}
	\end{subfigure}
	\begin{subfigure}{.33\textwidth}
		\centering
		\includegraphics[width=.86\linewidth]{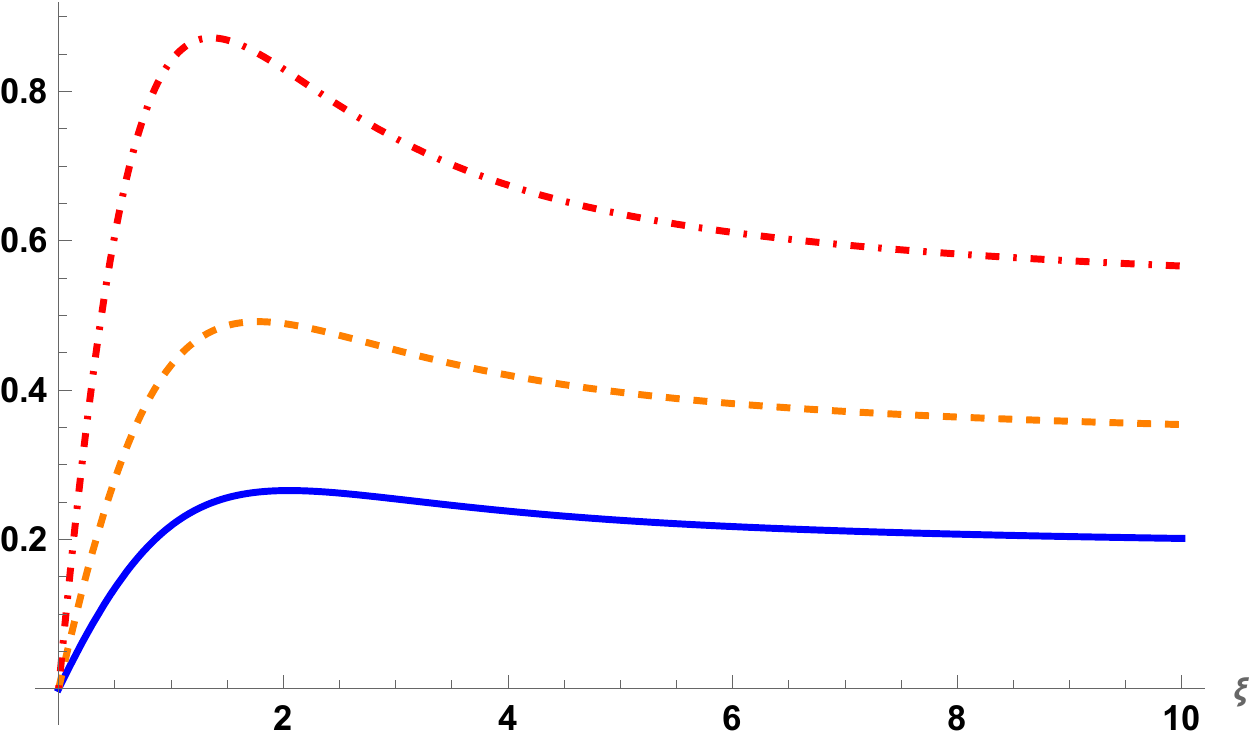}
		\caption{W-state. 
		}
		\label{fig:W_mono}
	\end{subfigure}
	\begin{subfigure}{.33\textwidth}
		\centering
		\includegraphics[width=.86\linewidth]{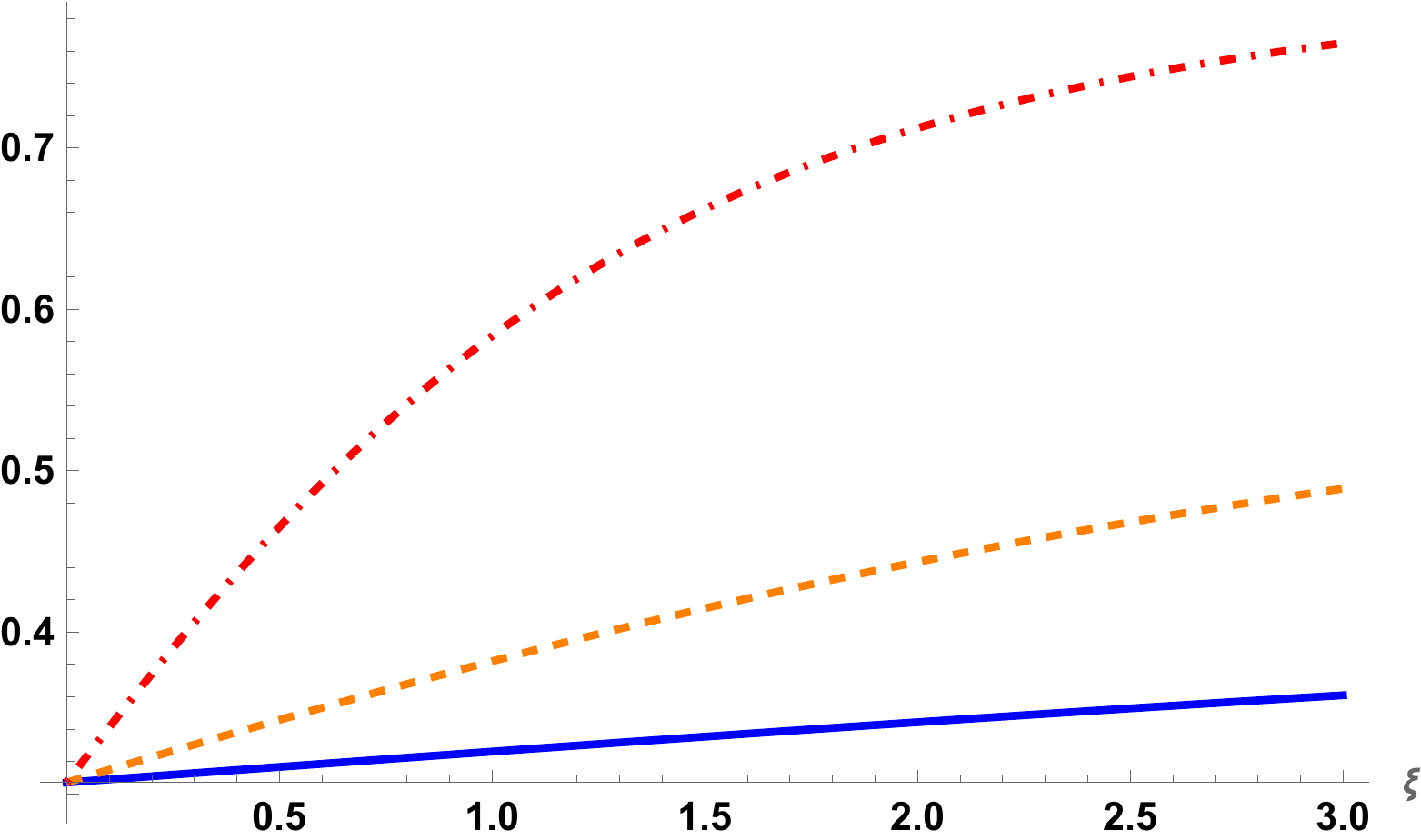}
		\caption{GHZ-state. 
		}
		\label{fig:W_mono}
	\end{subfigure}
	\caption{  $S^{(\xi)}_R(A:B\bar{B})-S^{(\xi)}_R(A:B)$ are plotted as a function of Renyi index $\xi$ for $r=\frac{\pi}{16}$ (solid blue), $r=\frac{\pi}{8}$ (orange dashed) and $r=\frac{\pi}{4}$ (red dot-dashed). }
	\label{mono}
\end{figure}

In this section we show some representative plots of the monotonicity of the reflected entropy by depicting  $S^{(\xi)}_R(A:B\bar{B})-S^{(\xi)}_R(A:B)$ as a function of Renyi index $\xi$ for Bell, Werner and GHZ states. We show that $S^{(\xi)}_R(A:B\bar{B})-S^{(\xi)}_R(A:B)$ is always positive for any value of $\xi$ which indicates that reflected entropy (Renyi index $\xi=1$) is a valid correlation measure for the systems under question. We have considered all the possible configurations of the parties to check the monotonicity where in \cref{mono} only three representatives have been presented.

Nevertheless, we may elaborate more on the discussion of the main text regarding the existence in general of other quantum states violating  monotonicity of the reflected entropy under partial trace we generalize them even for higher dimensional Hilbert spaces. The violation depends on the ratio $p=\frac{a}{b}$ which changes with the dimension of the Hilbert space. Such a generic state in  $\mathcal{H}_A\otimes\mathcal{H}_B\otimes\mathcal{H}_C=\mathbb{C}^{n+1}\otimes\mathbb{C}^{m+1}\otimes\mathbb{C}^2$ can be suggested to be
\begin{equation}
\begin{aligned}
			\rho_{A B C}= \frac{1}{2 n a+ 2(m-1)b}\Big[a|000\rangle\langle 000|+a| 110\rangle\langle 110|+\sum_{m,n}\Big(&a| n00\rangle\langle n00|+a| n10\rangle\langle n10|) \\ &+b|0m0\rangle\langle 0m0|+b|1m1\rangle\langle 1m1|\Big)\Big]~,
\end{aligned}\label{gen_prob}
\end{equation}
where $n,m\geq2$. Considering $n=m=2$, we get the states given in \cite{Hayden:2023yij}. The state presented in \cref{rhoABC} can be reproduced by taking $n=3$ and $m=2$ in \cref{gen_prob}. We expect that for any arbitrary values of $m$ and $n$, the plots for $S^{\xi}_{R}(A:BC)-S^{\xi}_{R}(A:B)$ with respect to the Renyi index and $p$ are similar to those presented in \cref{fig_problem,fig_problem_2}. The generic state in \cref{gen_prob} represents the class of states showing the non-monotonicity of the reflected entropy. It would be interesting to study the characteristics of these states in detail compared to the states that respect the monotonicity under partial tracing.

\end{appendices}


\bibliographystyle{JHEP}


\providecommand{\href}[2]{#2}\begingroup\raggedright\endgroup

\end{document}